\documentclass[12pt]{article}
\usepackage{latexsym}
\begin{document}
\thispagestyle{empty}

\begin{center}{\bf HEAT KERNEL EXPANSION\\
IN THE COVARIANT PERTURBATION THEORY}
\end{center}
\vspace*{0.37truein}
\begin{center}
Yuri V. Gusev\\[2mm]
IRMACS Centre, Simon Fraser University,\\
Burnaby, BC, Canada V5A 1S6\\
e-mail: yu\_gusev@yahoo.com\\
\end{center}

\vspace*{0.21truein}

\centerline{\bf Abstract}

{\noindent
Working within the framework of the covariant perturbation theory, we obtain the coincidence limit of the heat kernel of an elliptic second order differential operator that is applicable to a large class of quantum field theories. The basis of tensor invariants of the curvatures of a gravity and gauge field background, to the second order, is derived, and the form factors acting on them are obtained in two integral representations. The results are verified by the functional trace operation,  by the short proper time (Schwinger-DeWitt) expansions, as well as  by the computation of the Green function for the two-dimensional scalar field model.}
\textwidth=6.0 in
\textheight=9.0 in
\vsize=9.0 in
\hsize=6.0 in
\voffset=-23mm
\hoffset=-15mm
\\
\\
\\
\linewidth=6.0 in
\pagebreak

\section{Introduction}

The heat kernel is a unique mathematical subject that represents a fundamental block for building any quantum field theory \cite{Vilk-Strasb}. This subject was extensively studied in mathematics \cite{Gilkey-book} and in physics \cite{Avramidi-book}. For the general review of the current state of research we recommend  \cite{Vassilevich-PhysRep}. The present paper is a technical but important step in the development of one of the directions in heat kernel research, the {\em covariant perturbation theory (CPT)} proposed by A. Barvinsky and G. Vilkovisky \cite{CPT1}. In this paper we use notations and techniques of the preceding papers on the covariant perturbation theory \cite{CPT2,CPT3,CPT4}, however, to keep this work self-contained we include their basic formulas. This may be needed because the bulk of \cite{CPT4,CPT-book} is not published in journals or electronically.

The setup and algorithms of CPT were outlined in \cite{CPT2}, whose main ideas are briefly repeated in this Introduction and at the beginning of the section~\ref{cpt}. We consider quantum field theories in space-times with the metric tensors of constant positive signature, i.e., Euclidean space-times.\footnote{The metric $g^{\mu\nu}(x)$ is
 characterized by  its Riemann and Ricci curvatures
$R^\mu_{\,\cdot\, \alpha\nu\beta}=\partial_\nu
\Gamma^\mu_{\alpha\beta} - \ldots,\ \ 
R_{\alpha\beta}= R^\mu_{\,\cdot\, \alpha\mu\beta}$.
} 
The space-time dimension $2\omega$ is
arbitrary throughout the paper, except for the section \ref{Weyl}.
If $S[\varphi]$ is a classical action of a field theory, then the induced differential operator is,
\begin{equation}
F_{AB}(\nabla)\delta(x,y)=
\frac{\delta}{\delta \varphi^A(x)}
\frac{\delta}{\delta \varphi^B(y)} S[\varphi]. \label{hessian}
\end{equation}
The operator ${F}(\nabla)$ acts on small fluctuations of an arbitrary set of fields $\varphi^A(x)$. Throughout the paper, like in any other work on the covariant perturbation theory, we employ DeWitt's condensed notations \cite{DeWitt-book65}.
In this convention the field index $A$ stands for any set of discrete
indices of tensor-spinor fields which allows us
to describe a wide range of field models from scalar 
to non-Abelian gauge fields. That is it, all results presented here are valid for arbitrary background non-Abelian fields. Also, in DeWitt's notations the integrals over the spacetime are omitted except for the overall integral.

The heat kernel $\hat{K}(s|x,y)$ is assumed to be a solution of the heat equation in the form,
\begin{equation}
\frac{\partial }{\partial s}\hat{K}(s|x,y)=
\hat{F}(\nabla^x)\hat{K}(s|x,y),     \label{heateq}
\end{equation}
where $s$ is the proper time \cite{Schwin-PR51,DeWitt-book65,DeWitt-book03}  along the geodesic connecting two space-time points $x$ and $y$,
and operator $\hat{F}(\nabla)$ is acting on fields at point $x$.
The hat notation in (\ref{heateq}) and elsewhere indicates matrix valued quantities. The matrix conventions are
$\hat{1}=\delta^A{}_B,\hspace{7mm} \hat{P}=P^A{}_B,$ etc.
The matrix trace over the index set $A$ is denoted by ${\rm tr}$, e.g.,
${\rm tr}\hat{P}=P^A{}_A$.
We restrict ourselves to quantum field models that can be characterized by
the elliptic second order differential operator of the type called in physics literature
a minimal second order operator \cite{BarVilk-PRep85},
\begin{equation}
\hat{F}(\nabla)=
\Box \hat{1}+ \hat{P}- \frac16R\hat{1},  \label{intro-2}
\end{equation}
where the Laplace-Beltrami operator (or Laplacian),
\begin{equation}
\Box \equiv g^{\mu\nu}\nabla_\mu\nabla_\nu,   \label{box}
\end{equation}
is constructed in terms of the covariant derivative $\nabla_{\mu}$,
which is specified by its commutator curvature
\begin{equation}
[\nabla_\mu, \nabla_\nu]\varphi^A\equiv
(\nabla_\mu\nabla_\nu-\nabla_\nu\nabla_\mu)\varphi^A =
{\cal R}^A{}_{B\mu\nu}\varphi^B. \label{intro-5}
\end{equation}
The explicit presence of the Ricci scalar term in the operator (\ref{intro-2}) rather than in the potential is related to conformal models studies in four dimensions \cite{CPT2}.
The set of the field strengths
\begin{equation}
R^{\alpha\beta\mu\nu},\hspace{7mm} \hat{\cal R}_{\mu\nu},
\hspace{7mm} \hat{P} \label{intro-6}
\end{equation}
characterizing the background will be called the curvatures denoted by the symbol $\Re$. Manifolds under consideration are asymptotically
flat with trivial topology, specifically, their gauge and gravitational
curvatures and potential (\ref{intro-6}) vanish at the space-time infinity \cite{CPT2}.
The work \cite{BarVilk-PRep85} contains a detailed review of the classes of models which are associated with the operator (\ref{intro-2}) and, therefore, can be dealt with by the present formalism.

The formal operator definition of the heat kernel is
\begin{equation}
\hat{K}(s|x,y)=
	   \exp \Big[ s \hat{F}(\nabla^x) \Big]
	    \delta(x,y). \label{intro-1}
\end{equation}
Knowledge of the heat kernel with separated points $\hat{K}(s|x,y)$ would allow one to construct the covariant diagrammatic expansion
to all loop orders 
\cite{DeWitt-book65,BarVilk-QFTQS87}. Here we focus only on the coincidence limit of the heat kernel $\hat{K}(s|x,x)$. As a result, quantum field theory applications are limited to the one-loop approximation. In particular, the one-loop effective action is defined as \cite{BarVilk-PRep85},
\begin{equation}
W=\frac12 {\rm Tr}\ {\rm ln} F
- \int d^{2\omega}x\delta^{(2\omega)}(x,x)(\dots), \label{Trln}
\end{equation}
and it represents the generating function for the one-particle irreducible diagrams \cite{DeWitt-book65,Vasiliev-book98}.
This fact is especially important for quantum gravity
where the work with ordinary Feynman diagrams would be formidable.
The second term in (\ref{Trln}) stands for the contribution of the local functional measure \cite{FradVilk-PRD73}. As shown in \cite{FradVilk-LNC77} this contribution always cancels the volume divergences. For the massless operators (\ref{intro-2}) the result of such cancellation is a subtraction of the zero curvature term in the heat kernel expansion.  Then, the one-loop effective action $W$ is given by 
the trace of the heat kernel \cite{CPT2},
\begin{equation}
-W=\frac12\int^\infty_0 \frac{ds}s
\left({\rm Tr} K(s) - {\rm Tr} K(s)|_{\Re=0}\right), \label{intro-7}
\end{equation}
where ${\rm Tr}$ denotes the functional trace
\begin{equation}
{\rm Tr} K(s)=\int d^{2\omega}x\,
{\rm tr} \, \hat{K}(s|x,x). \label{intro-8}
\end{equation}
From this point, we do not indicate the space-time dimension at the integral measure, i.e., $d\, x \equiv d^{2\omega}\, x$.
The knowledge of the heat kernel, in contrast to its trace (\ref{intro-8}),
would provide one with the Green function or propagator 
$\hat{G}(x,y)$ defined by
\begin{equation}
\hat{F}(\nabla^x) \hat{G}(x,y)=
	    - \hat{1}\delta(x,y), \label{propagator}
\end{equation}
due to the Schwinger equation,
\begin{equation}
\hat{G}(x,y)=  \int_0^{\infty} \! d\, s \hat{K}(s|x,y).  \label{Schwinger}
\end{equation}

The ultraviolet divergences appear in quantum field theories as
divergences of the proper time integrals (\ref{intro-7}), (\ref{Schwinger})
at the lower limit. They are to be removed by the renormalization
procedure \cite{DeWitt-book65,BarVilk-PRep85}.
For massive field theories, there is a mass factor ${\rm e}^{-s m^2}$
in the heat kernel (trace), which makes the integral converge at the upper limit $s \rightarrow \infty$.
Early techniques (Schwinger-DeWitt expansion) could only reproduce the asymptotic behaviour 
of the heat kernel $K(s)$ at $s \rightarrow 0$. As a result, these proper time integrals for massless field theories would diverge at the upper limit.
These infrared divergences due to the method of computation were irrelevant of a quantum field theory under study. This is clearly seen from the fact that, upon the proper time integration,
the short time expansion of the heat kernel corresponds to the inverse large mass expansion
of the effective action or the Green functions \cite{DeWitt-book65}.
Therefore, a method that would allow one
to compute the heat kernel in the whole range of the proper time values was needed. 
Such a method, proposed by G.A. Vilkovisky  \cite{Vilk-Gospel} and called the covariant perturbation theory,
was systematically developed in a series of works 
\cite{CPT1,CPT2,CPT3,CPT-book}.
CPT is  the covariant expansion of the heat kernel by orders of the curvatures. 
The CPT heat kernel contains an infinite number of derivatives, expressed as form factors, acting on the background field curvatures. Thus, it is a nonlocal expression, and the heat kernel form factors turn into the Green functions acting on the curvatures.
For the heat kernel trace up to the third order in the curvatures it has the general form 
\cite{BGVZ-JMP94-bas,BGVZ-JMP94-asymp} (a different method leading to the nonlocal effective action was developed in \cite{Avram-PLB90}):
\begin{eqnarray}
{\rm Tr}\, K(s)&=&
\frac{1}{(4 \pi s)^{\omega}} \int \! d\, x\, g^{1/2}(x)
{\rm tr} \left\{ {1} + s \Re
+ s^2 \sum_{i=1}^{5} f_{i}(s, \Box_1, \Box_2) \Re_1 \Re_2(i) \right.
\nonumber\\&&\left. \mbox{}
	+ s^3 \sum_{i=1}^{29} F_{i}(s, \Box_1, \Box_2 , \Box_3 ) 
	\Re_1 \Re_2 \Re_3(i)
	+ {\rm O}[\Re^4] \right\},  \label{TrKsymb}
\end{eqnarray}
where $g$ is the determinant of the metric tensor, and $f_i $ and $F_i$ are for analytic functions (form factors) of the dimensionless quantities
\begin{equation}
\xi=-s\Box, \label{xi}
\end{equation}
which act on tensor invariants constructed of the curvatures $\Re$. 
The index of an operator $\Box$ indicates the local curvature it is acting on, i.e.,
$ \Box_2 R_1 \hat{P}_2\equiv R(x) (\Box \hat{P} (x))$. The expressions resulting from these operations, in the curly brackets in (\ref{TrKsymb}), are taken at the same spacetime integration point $x$.

The CPT calculations are carried out with accuracy ${\rm O}[\Re^n]$, meaning the derived expressions contain terms of up to $n-1$ order in the curvatures explicitly. 
In contrast to the short proper time (Schwinger-DeWitt) expansion of the heat kernel,
the expression (\ref{TrKsymb}) is valid for any value of the proper time $s$.
In general, as is shown in \cite{CPT2},  the large time behavior of ${\rm Tr} K(s)$ is
\begin{equation}
{\rm Tr} K(s) \propto s^{1-\omega}, \ \ \ s \rightarrow \infty, \ \ \ 
\Re \neq 0, \label{trklarges}
\end{equation}
for all curvature orders, except the zeroth.
This property guarantees the convergence of field theory integrals
at the upper limit in space-time dimensions $2\omega >2$.

In this work we derive the coincidence limit of the heat kernel (\ref{intro-1})  up to the second order in curvatures. Even though the final result is the same, this task is accomplished by two different methods. One method is a direct application of the CPT formulas \cite{CPT2}.
The other is the generating function method. The method of generating expressions for the Green functions first has been proposed for analysis of local divergences of the coincidence
limits of the Green  functions \cite{BarVilk-QFTQS87,Ball-PRep89}.
It reflects the fundamental feature of the effective action
as the generating function of the one particle irreducible
Green functions. In \cite{BarGus-CQG92} it was shown  that this method can be used to treat nonlocal curvature expansions as well.
The work with heat kernels, instead of the Green functions, has the advantage that one is not restricted to specific space-time dimensions, and no
divergences are present until the proper time integrals are done.
Since the differential operator $F(\nabla)$ 
depends on three independent background fields, 
the metric, the gauge field, and the potential,
there are three possible variational equations
\cite{BarVilk-QFTQS87,Ball-PRep89,BarGus-CQG92}.
The one we are interested in is with respect to the potential $\hat{P}$,
\begin{equation}
\hat{K}(s|x,x)=
\frac{1}{s}\frac{\delta}{\delta \hat{P}(x)}
{\rm Tr} {K}(s).                \label{varP}
\end{equation}

We start from the trace of the heat kernel (\ref{TrKsymb}). The only important feature of the form factors in (\ref{TrKsymb}) required here is that they are functions of the operator (\ref{xi})
and do not depend explicitly on the curvatures $\Re$. This is the feature of the variational method which always eliminates one curvature reducing an accuracy ${\rm O}[\Re^n]$ by one order. In this case, the variation of form factors is not required.
By this approach we can only obtain the terms valid to the second order
in the curvatures of the form \cite{BarGus-Winn94}:
\begin{equation}
\hat{K}(s|x,x)= \frac{g^{1/2}(x)}{(4 \pi s)^{\omega}}
\left\{ s \sum_{i=1}^{2} g_i(s, \Box) \Re[i] +
	s^2 \sum_{i=1}^{11} G_i(s, \Box_1, \Box_2, \Box_3) 
\Re_1 \Re_2[i]
	+ {\rm O}[\Re^3] \right\}. \label{Ksymb}
\end{equation}
In the next section, we also derive this expression directly by the CPT algorithms
and confirm the equivalence of two methods.

An important element of the present study is use of computer symbolic manipulations. There are two different kinds of symbolic manipulations performed.
One is purely algebraic manipulations of form factors which
was performed with general purpose mathematics software {\em Maple}.
The other is manipulations with tensors in order to analyze tensor invariants, but especially
to compare these results with the Schwinger-DeWitt coefficients from other works.
 These manipulations were performed with help of {\em MathTensor} \cite{MathTensor} and {\em Ricci} \cite{Ricci} packages, both of which work under {\em Mathematica} software. 

This paper represents a natural part of the series of works on the covariant
perturbation theory \cite{CPT1,CPT2,CPT3,CPT4,CPT-book}, and it consists of six sections. Sections \ref{cpt} to \ref{explicitK2} contain main results, namely,  the heat kernel coincidence limit up to the second order in the curvatures.  We start from the basic perturbations expansion and proceed
to the covariant expressions. Its form factors are obtained
in two integral representations.
In section \ref{shortK2} we study the short proper time 
asymptotic behaviour of $K(s)$ and compare these asymptotics with 
the known Schwinger-DeWitt coefficients; the large (late) time asymptotic is reproduced as well.  Section \ref{Weyl} is different from others, since it deals with a particular class of field models,  the Weyl invariant model in two dimensions.
We compute the Green function for this model and show that
the result coincides with the known closed form solution. 
In Summary we address possible applications of the obtained
results for QFT models. 


\section{Covariant perturbation theory for the heat kernel} \label{cpt}

Direct computations according to the covariant perturbation theory 
rely on \cite{CPT2} whose basic equations we reproduce here to make the present 
paper self-contained.
We also emphasize that the work with form factors of the second order of the heat kernel is very much similar
to the work with the third order in the heat kernel trace.
Therefore, in this and the following sections we use notations and techniques developed in \cite{CPT3,CPT4,CPT-book}.

To set up the perturbation theory,
splitting of the metric and the covariant
derivative into auxiliary parts and perturbations is introduced:
\begin{eqnarray}
g^{\mu\nu}&=&\widetilde{g}^{\mu\nu} + h^{\mu\nu}, \label{tildeg}
\\
\nabla_{\mu}\varphi &=& \widetilde{\nabla}_{\mu}\varphi
	+ \hat{\Gamma}_{\mu}\varphi.   \label{tildeD}
\end{eqnarray}
The auxiliary metric and derivative are taken to be flat, i.e.,
\begin{eqnarray}
&& R_{\mu\nu\alpha\beta}(\widetilde{g})=0,  \label{zeroR}
\\
&&[
\widetilde{\nabla}_{\mu},\widetilde{\nabla}_{\nu} 
] \varphi=
\widetilde{\hat{\cal R}}_{\mu\nu} \varphi = 0 .
\label{zeroRe}
\end{eqnarray}

Now there are three independent perturbations,
one each for the metric, the connection and the potential
\begin{equation}
h^{\mu\nu},\hspace{7mm} \widehat{\Gamma}_\mu,\hspace{7mm}
\hat{P}-\frac16R\hat{1}.                      \label{third-2}
\end{equation}
In terms of these perturbations
the differential operator (\ref{intro-2}) is divided
as follows,
\begin{equation}
\hat{F}(\nabla)=\widetilde{\vphantom{I}\Box} + V(\widetilde{\nabla}),
\end{equation}
with the flat Laplacian
\begin{equation}
\widetilde{\vphantom{I}\Box}=\widetilde{g}^{\mu\nu}
\widetilde{\nabla}_\mu\widetilde{\nabla}_\nu
\end{equation} 
and the perturbation
\begin{eqnarray}
V(\widetilde{\nabla}) &=& h^{\mu\nu} \widetilde{\nabla}_{\mu}
\widetilde{\nabla}_{\nu} + 2 \hat{\Gamma}^{\mu} \widetilde{\nabla}_{\mu}
+\hat{P} - \frac16 R \hat{1},  \label{pertV}
\end{eqnarray}
where
\begin{equation}
\hat{\Gamma}^{\mu} \equiv(\widetilde{g}^{\mu\nu}
+h^{\mu\nu})\hat{\Gamma}_{\nu} \label{third-11}
\end{equation}
(otherwise, raising and lowering indices are done
with help of the auxiliary metric $\widetilde{g}^{\mu\nu}$).

The heat kernel is expanded then in powers of
the perturbation \cite{Schwin-PR51,CPT2}:
\begin{equation}
K(s) = \sum^\infty_{n=0} K_n(s)   \label{third-1}
\end{equation}
where $K_n(s)$ is a term of the $n$-th power in the perturbations
(\ref{third-2}).
Perturbative solution to (\ref{heateq}) can be found iteratively
and has the general form,
\begin{equation}
K_n(s) = \int_0^s d t_n \int_0^{t_n} d t_{n-1} \ldots  \int_0^{t_2} d t_{1}
	{\rm e}^{(s-t_n)\widetilde{\vphantom{I}\Box}} V
	{\rm e}^{(t_n- t_{n-1})\widetilde{\vphantom{I}\Box}} V
	\ldots
	{\rm e}^{(t_2-t_1)\widetilde{\vphantom{I}\Box}} V
	{\rm e}^{t_1\widetilde{\vphantom{I}\Box}},
\label{dayson}
\end{equation}
which is in fact merely the Dyson series 
\cite{Schwin-PR51,Dayson-PR49}.
For the simple case of flat space-time without background 
gauge fields, the perturbation is just the potential term $\hat{P}$.
In this case a closed form
of $K_n(s)$ for any $n$ can be easily written down
\cite{WilkFujOsb-PRA81}.

An exact solution for the zeroth order of the heat kernel
is known
\cite{DeWitt-book65},
\begin{equation}
K_0(s|x,y)= \frac1{(4 \pi s)^{\omega}}\widetilde{g}^{1/4}(x)
	\widetilde{g}^{1/4}(y)
	{\rm e}^{-\frac{\widetilde{\sigma}(x,y)}{2s}}
	\widetilde{a}_0(x,y).   \label{K0}
\end{equation}
Here $\widetilde{a}_0$ is the parallel transport operator along
the geodesic connecting $y$ to $x$, 
$\widetilde{g}$ is the determinant
of the auxiliary metric $\widetilde{g}^{\mu\nu}$, and 
$\widetilde{\sigma}$
is the auxiliary world function (in the Cartesian coordinates it is
simply $(x-y)^2 /2$).
When calculated by the algorithms of  \cite{CPT2},
term $K_n(s)$ is obtained in the generic form,
\begin{eqnarray}
&& \hat{K}_n(s|x,x) = \frac1{(4\pi s)^\omega}
\widetilde{g}^{1/2}(x)\,
\int_{\alpha_i\geq0}d^n\alpha\,
  \delta(1-\sum^n_1\alpha_i)   \nonumber\\&&
\times\exp\Big[s\Omega_{n+1}
  (\alpha_1,\dots\alpha_{n+1}|\tilde{\nabla}^i)\Big]
\sum^{n}_{l=0}s^l\widehat{C}^l_n                   \label{Kser}
(\alpha_1,\dots\alpha_{n+1}|x_i)\Big|_{x_i=x}.
\end{eqnarray}
With the notation
\begin{equation}
\int_{\alpha_i\geq0}d^n\alpha\,
  \delta(1-\sum^n_1\alpha_i)
  f(\alpha_1,\dots\alpha_n|x_i)\Big|_{x_i=x}
= \big<f\big>_n,                                 \label{third-4}
\end{equation}
this can be rewritten
\begin{equation}
\hat{K}_n(s) = \frac1{(4\pi s)^\omega}
\widetilde{g}^{1/2}\,
\sum^{n}_{l=0}s^l\big<
{\rm e}^{s\Omega_{n+1}}
\widehat{C}^l_n\big>_n.
\end{equation}
Here $\Omega_n(\alpha_1,\dots\alpha_n|{\widetilde{\nabla}}^i)$ is an
operator of the second order in ${\widetilde{\nabla}}^i$, and
the derivatives ${\widetilde{\nabla}}^i$ act on the perturbation with 
index $i$
contained in ${\hat{C}}^l_n$. Each term in
$\widehat{C}^l_n(\alpha_1,\dots\alpha_n|x_i)$,
where $i=1,\ldots n$, is a product
of $n$ perturbations (\ref{third-2}) at the points
$x_1,\dots x_n$ respectively, and the index
 $i$ 
on a perturbation means that the perturbation
is at the point $x_i$, e.g, $\hat{P}_1=\hat{P}(x_1)$.
After the action of ${\widetilde{\nabla}}^i$ is done, 
all points $x_i$
are made coincident with the point $x$ in (\ref{Kser}). 
For the consistency with \cite{CPT3,CPT4}, 
in the $n$-th order, we use the notation
$\Omega_{n+1}$ instead of $\Omega_n$.

Let us now display the forms for 
$\Omega$ and $\hat{C}$, for $n=0$,
\begin{eqnarray}
\Omega_1 &=& 0,\\
\hat{C}^0_0 &=& \hat{1}.
\end{eqnarray}
The results for $n=1$ are
\begin{eqnarray}
\Omega_2(\alpha_1,\alpha_2|\tilde{\nabla}^i)
  &=& \alpha_1\alpha_2\widetilde{\vphantom{I}\Box}, \label{omega2}\\
\hat{C}^0_1(\alpha_1,\alpha_2|x_i)& =&
\frac12h \hat{1},\\
\hat{C}^1_1(\alpha_1,\alpha_2|x_i) &=&
  \alpha_1^2
    \Big(\tilde{\nabla}_\mu\tilde{\nabla}_\nu h^{\mu\nu}
-\frac16 R\Big)\hat{1}
 +\hat{P}
  -2\alpha_1 \tilde{\nabla}_\mu\hat{\Gamma}^\mu,
\end{eqnarray}
and for $n=2$
\begin{eqnarray}
\Omega_3(\alpha_1,\alpha_2,\alpha_3|\tilde{\nabla}^i)& =&
  \alpha_2\alpha_3\tilde{\vphantom{I}\Box}_1
  +\alpha_1\alpha_3\tilde{\vphantom{I}\Box}_2
  +\alpha_1\alpha_2\tilde{\vphantom{I}\Box}_3, \label{omega3}
\\
\hat{C}^0_2(\alpha_1,\alpha_2,\alpha_3|x_i) &=&
  \Big(\frac14 h_1 h_2
+\frac12 h_{1}^{\mu\nu} h_{2}^{\alpha\beta}
\tilde{g}_{\alpha\mu}\tilde{g}_{\beta\nu}
\Big)\hat{1},
\\
\hat{C}^1_2(\alpha_1,\alpha_2,\alpha_3|x_i)& =&
-\hat{P}_1 h_2
-2 \tilde{g}_{\alpha\beta}\hat{\Gamma}_1^{\alpha}\hat{\Gamma}_2^{\beta}
\nonumber\\&&\mbox{}
+\Big(\tilde{g}_{\mu\nu}(D^3_{\alpha}+D^2_{\alpha})+
2\tilde{g}_{\mu\alpha}(D^1_{\nu}-D^3_{\nu})\Big)
\hat{\Gamma}_1^{\alpha}h_2^{\mu\nu}
\nonumber\\&&\mbox{}
+\hat{1}\left[\frac16 R_1 h_2
 -\frac12(
      \tilde{g}_{\mu\nu} D^1_\alpha D^1_\beta
     +\tilde{g}_{\alpha\beta} D^3_\mu D^3_\nu \right.
\nonumber\\&&\left.\mbox{}
     +4\tilde{g}_{\mu\alpha} D^3_\nu D^1_\beta )
      h_1^{\mu\nu} h_2^{\alpha\beta}\right],
\nonumber\\
\\
\hat{C}^2_2(\alpha_1,\alpha_2,\alpha_3|x_i) &=&
 \Big(\hat{P}_1-\frac16 R_1\hat{1}\Big)
\Big(\hat{P}_2-\frac16 R_2\hat{1}\Big)
\nonumber\\&&\mbox{}
  +(D^1_\alpha D^1_\beta +D^3_\alpha D^3_\beta)
\Big(\hat{P}_1-\frac16 R_1\hat{1}\Big)h_2^{\alpha\beta}
\nonumber\\&&\mbox{}
  -2( D^1_\mu+D^3_\mu)
\Big(\hat{P}_1-\frac16 R_1\hat{1}\Big)\hat{\Gamma}^\mu_2
\nonumber\\&&\mbox{}
  -2( D^2_\alpha D^3_\mu D^3_\nu
     +D^3_\alpha D^1_\mu D^1_\nu )
\hat{\Gamma}^\alpha_1 h_2^{\mu\nu}
\nonumber\\&&\mbox{}
  +4 D^3_\alpha D^1_\beta \hat{\Gamma}_1^\alpha\hat{\Gamma}_2^\beta
  +D^3_\mu D^3_\nu D^1_\alpha D^1_\beta
     h_1^{\mu\nu} h_2^{\alpha\beta}\hat{1},
\end{eqnarray}
here
\begin{eqnarray}
h&=&h^{\mu\nu}\widetilde{g}_{\mu\nu},
\\
D^1_\mu &=& \alpha_3{\widetilde{\nabla}}^2_\mu
-\alpha_2{\widetilde{\nabla}}^3_\mu,  \label{third-21}
\\[\baselineskip]
D^2_\mu &=& \alpha_1{\widetilde{\nabla}}^3_\mu
-\alpha_3{\widetilde{\nabla}}^1_\mu,  \label{third-22}
\\[\baselineskip]
D^3_\mu &=& \alpha_2{\widetilde{\nabla}}^1_\mu
-\alpha_1{\widetilde{\nabla}}^2_\mu.  \label{third-23}
\end{eqnarray}
The variational method was applied directly
to the results for the trace of the heat kernel
in terms of perturbations derived in \cite{CPT4},
and it resulted in the same expression above. 

Finally, one should make the series (\ref{Kser}) manifestly covariant.
This means we need to replace the perturbations
and the auxiliary metric and derivative (\ref{tildeg})--(\ref{tildeD})
by the respective covariant curvatures (\ref{intro-6}), 
metric and derivative (\ref{intro-5}). A covariant expansion 
to the second order in the curvatures involves the first order in the
perturbations as well.
The curvature expansions of the perturbations 
(\ref{third-2}) can be obtained
from eqs. (\ref{zeroR})--(\ref{zeroRe}).
Their general solutions are integral
equations which were solved by iterations in terms
of $R_{\mu\nu}$ and ${\cal R}_{\mu\nu}$ in 
 \cite{CPT2}, Eqs.~(4.28), (4.29).
Alternatively, one can use the generating function method (\ref{varP})
applied to ${\rm Tr}K(s)$ of (\ref{TrKsymb}).
It is easy to observe that only thirteen of the 29 tensor structures
of the trace of the heat kernel presented in \cite{BGVZ-JMP94-bas}
contain matrix $\hat{P}$ and thus contribute
to the heat kernel itself.

Whether we obtain the covariant heat kernel 
by the generating function method
or by direct computations,
the result is
\begin{eqnarray}
\hat{K}(s|x,x)&=&\frac{1}{(4\pi s)^{\omega}}
g^{1/2}
\Big\{\hat{1}
+s\left(
g_1(-s\Box)\hat{P}
+g_2(-s\Box)R\hat{1}\right)
\nonumber\\&&
+s^2 \sum_{i=1}^5 G_i(-s\Box_1,-s\Box_2,-s\Box_3) \Re_1 \Re_2[i]
\nonumber\\&&
+s^3\sum_{i=6}^{11} G_i(-s\Box_1,-s\Box_2,-s\Box_3)
\Re_1\Re_2[i]
\nonumber\\&&\mbox{}
+s^3 G_{12}(-s\Box_1,-s\Box_2,-s\Box_3)
\Re_1\Re_2[12]
+\rm{O}[\Re^3]\Big\}.         \label{K2}
\end{eqnarray}
Here $\Re_1\Re_2[i]$ with $i=1$ to $12$ are quadratic structures:
\begin{eqnarray}
&&\Re_1 \Re_2 [1]= \hat{P}_1 \hat{P}_2, \label{RRK1}
\\
[\baselineskip]
&&
\Re_1 \Re_2 [2]=  \hat{\cal R}^{\mu\nu}_1 \hat{\cal R}_{2 \mu \nu},\label{RRK2}
\\
[\baselineskip]
&&
\Re_1 \Re_2 [3]= \hat{P}_1 R_2,\label{RRK3}
\\
[\baselineskip]
&&
\Re_1 \Re_2 [4]= R_1 R_2 \hat{1},\label{RRK4}
\\
[\baselineskip]
&&
\Re_1 \Re_2 [5]= R^{\mu\nu}_1 R_{2\mu\nu}\hat{1},\label{RRK5}
\\
[\baselineskip]
&&
\Re_1 \Re_2 [6]=
\nabla_{\mu}\hat{\cal R}^{\mu\nu}_1
\nabla^{\alpha}\hat{\cal R}_{2\alpha\nu},\label{RRK6}
\\
[\baselineskip]
&&
\Re_1 \Re_2 [7]=
\big[\nabla_{\alpha}\hat{P}_1,
\nabla_{\beta}\hat{\cal R}_2^{\beta\alpha}\big],\label{RRK7}
\\
[\baselineskip]
&&
\Re_1 \Re_2[8]= \nabla_\mu \nabla_\nu \hat{P}_1 R_2^{\mu\nu},\label{RRK8}
\\
[\baselineskip]
&&
\Re_1 \Re_2 [9]=\nabla_\alpha R_{1\mu\nu}
\nabla^{\mu}R_2^{\nu\alpha}\hat{1},\label{RRK9}
\\
[\baselineskip]
&&
\Re_1 \Re_2 [10]=\nabla_\mu \nabla_\nu R_1 R_2^{\mu\nu}\hat{1},\label{RRK10}
\\
[\baselineskip]
&&
\Re_1 \Re_2 [11]= \nabla_\alpha \nabla_\beta R_{1\mu\nu}
\nabla^\mu \nabla^\nu R_2^{\alpha\beta}\hat{1}.           \label{RRK11}
\end{eqnarray}
There is an additional quadratic structure
linear in $\hat{{\cal R}}_{\mu\nu}$
\begin{equation}
\Re_1\Re_2[12]=\nabla_{\nu}\hat{{\cal R}}_1^{\nu\mu}\nabla_{\mu}R_2,
\label{RRK12}
\end{equation}
which is separated from the others because
it is absent in the final answer.
Without gravity, the basis (\ref{RRK1})--(\ref{RRK12})
reduces to only four non-vanishing curvature structures.


\section{Form factors of the nonlocal heat kernel} 
\label{explicitK2}

The form factors
$g_i (-s \Box)$
and $G_i(-s\Box_1,-s\Box_2,-s\Box_3)$
are obtained as the integrals over parameters (\ref{third-4})
and, in this form, represented by two nonlocal kernels
(cf., (\ref{omega2}) and (\ref{omega3})):
\begin{equation}
\exp (s\alpha_1\alpha_2\Box)  \label{secondff}
\end{equation}
and
\begin{equation}
\exp (s\Omega), \hspace{5mm}
\Omega=\alpha_2 \alpha_3 \Box_1 + \alpha_1 \alpha_3 \Box_2
 + \alpha_1 \alpha_2 \Box_3.  \label{thirdff} 
\end{equation}
The function (\ref{secondff}) appears in the combinations
\begin{eqnarray}
{\cal A}_{m} &=& {\rm e}^{s \alpha_1\alpha_2\Box_{m}}, 
\label{expan-12}
\\[\baselineskip]
{\cal B}_{m} &=& \frac{{\rm e}^{s \alpha_1\alpha_2\Box_{m}}-1}{s\Box_m}, 
\label{expan-13} \label{expan-14}
\\[\baselineskip]
{\cal U}_{mn} &=&
\frac{{\rm e}^{s \alpha_1\alpha_2\Box_{m}}-
{\rm e}^{s \alpha_1\alpha_2\Box_{n}}}{s(\Box_m-\Box_n)}, 
\label{expan-15}
\\[\baselineskip]
{\cal V}_{mn} &=& \frac1{s(\Box_m-\Box_n)}
\Big(\frac{{\rm e}^{s \alpha_1\alpha_2\Box_{m}}-1}{s\Box_m}-
\frac{{\rm e}^{s \alpha_1\alpha_2\Box_{n}}-1}{s\Box_n}
\Big), \label{expan-16}
\end{eqnarray}
where ($m,n=1,2,3; \ m\neq n$) and
the indices of the form factors refer to the indices
of the laplacians appearing in them.
The function (\ref{thirdff}) appears
in the combinations,
\begin{equation}
{\rm e}^{s\Omega},\hspace{7mm} 
{\rm e}^{s\Omega}-1. \label{alpha-12}
\end{equation}
The
coefficients of these functions are polynomials in $\alpha$~'s,
Laplacians, and inverse Laplacians (Green functions).
It is assumed that
$\Box_1$ acts on the curvature $\Re_1$,
$\Box_2$ acts on the curvature $\Re_2$,
and
\begin{equation}
\Box_3=(\nabla_1 + \nabla_2)^2.
\end{equation}
The operators $\Box_{m}$ of form factors commute
with the derivatives $\nabla{}$'s acting on the curvatures $\Re_m$ in
(\ref{RRK10})--(\ref{RRK11}), because contributions from such
commutations belong to the higher order ${\rm O}[\Re^3]$.

The first order form factors are formed of
the ones in (\ref{TrKsymb}) as,
\begin{eqnarray}
g_1(\xi)&=&2 f_4(\xi), \label{gf1}
\\[\baselineskip]
g_2(\xi)&=& f_3(\xi),\label{gf2}
\end{eqnarray}
here and below $\xi$ is the dimensionless operator-valued argument (\ref{xi}).
The second order form factors $G_i({\xi_1},{\xi_2},{\xi_3})$,
for $i=1$ to 11,
are expressed via form factors $F_i({\xi_1},{\xi_2},{\xi_3})$ of the trace
of the heat kernel (\ref{TrKsymb}), tabulated in  \cite{CPT4},
in the following way,
\begin{eqnarray}
G_1({\xi_1},{\xi_2},{\xi_3})&=&
F_1({\xi_1},{\xi_2},{\xi_3})
+F_1({\xi_2},{\xi_3},{\xi_1})
+F_1({\xi_3},{\xi_1},{\xi_2})
, \label{GF1}
\\
G_2({\xi_1},{\xi_2},{\xi_3})&=&
F_3({\xi_1},{\xi_2},{\xi_3}),\label{GF2}
\\
G_3({\xi_1},{\xi_2},{\xi_3})&=&
F_6({\xi_1},{\xi_3},{\xi_2})
+F_6({\xi_3},{\xi_1},{\xi_2})
-\frac12 ({\xi_3}-{\xi_1})
F_{17}({\xi_2},{\xi_3},{\xi_1})
,\label{GF3}
\\
G_4({\xi_1},{\xi_2},{\xi_3})&=&
F_4({\xi_1},{\xi_2},{\xi_3})+\frac14 ({\xi_3}-{\xi_2}-{\xi_1})
F_{15}({\xi_1},{\xi_2},{\xi_3}),\label{GF4}
\\
G_5({\xi_1},{\xi_2},{\xi_3})&=&
F_5({\xi_1},{\xi_2},{\xi_3}),\label{GF5}
\\
G_6({\xi_1},{\xi_2},{\xi_3})&=&
F_{14}({\xi_1},{\xi_2},{\xi_3}),\label{GF6}
\\
G_7({\xi_1},{\xi_2},{\xi_3})&=&
-F_{13}({\xi_2},{\xi_1},{\xi_3}),
\label{GF7}
\\
G_8({\xi_1},{\xi_2},{\xi_3})&=&
F_{17}({\xi_2},{\xi_1},{\xi_3})
+F_{17}({\xi_2},{\xi_3},{\xi_1})
,\label{GF8}
\\
G_9({\xi_1},{\xi_2},{\xi_3})&=&
F_{16}({\xi_1},{\xi_2},{\xi_3}),\label{GF9}
\\
G_{10}({\xi_1},{\xi_2},{\xi_3})&=&
-F_{15}({\xi_2},{\xi_1},{\xi_3}),
\label{GF10}
\\
G_{11}({\xi_1},{\xi_2},{\xi_3})&=&
F_{26}({\xi_1},{\xi_2},{\xi_3}), \label{GF11}
\\
G_{12}({\xi_1},{\xi_2},{\xi_3})&=&
F_{32}({\xi_3},{\xi_1},{\xi_2}),
\label{GF12}
\end{eqnarray}
where some interchanges and cyclic substitutions
of indices of the arguments $\xi_i$ are made.

It should be emphasized that these rules are not sensitive to
the representation of form factors.
When polynomials of $\alpha$~-parameters are present in (\ref{third-4}),
we call it $\alpha$~-representation.
In $\alpha$~-representation
the first order form factors are
\begin{eqnarray}
g_1&=&\left<{\cal A}\right>_{2}, \label{g1a}
\\
g_2&=&\left<\left({\alpha_1}^2-\frac16\right){\cal A}-{\cal B}\right>_{2}.
\label{g2a}
\end{eqnarray}

And the second order form factors admit the form
\begin{eqnarray}
G_1&=&\left< \rm{e}^{s\Omega_3}\right>_3, \label{G1a}
\\
G_2&=&\left< 2{\alpha_1}{\alpha_2}
\rm{e}^{s\Omega_3}\right>_3,\label{G2a}
\\
G_3 &=&
\left<
\frac12 \frac{1}{\Box_2}
\frac{{\cal A}_1}{s}
+\frac12 \frac{1}{\Box_2}
\frac{{\cal A}_3}{s}
+\frac12 \frac{\Box_1-\Box_3}{\Box_2}
{{\cal U}_{13}}
\right>_2
\nonumber\\&&\mbox{}
-\left<
2\frac{1}{\Box_2}
\frac{{\rm e}^{s\Omega_3}}{s}
+\left[
\Big(
\frac{2}{3}-2{\alpha_1}+2{\alpha_1}^2-{\alpha_2}+2{\alpha_1}{\alpha_2}\Big)
\right.\right.\nonumber\\&&\mbox{}\left.\left.
+\frac{\Box_1}{\Box_2}(2{\alpha_1}{\alpha_2}-{\alpha_2})
+\frac{\Box_3}{\Box_2}({\alpha_2}-2{\alpha_1}{\alpha_2})
\right]
{\rm e}^{s\Omega_3}
\right>_3,\label{G3a}
\\
G_4&=&
\left<
\frac{1}{\Box_1}({{\alpha_1}}^2-\frac16)\frac{{\cal A}_2}{s}
-\frac{1}{\Box_1}\frac{{\cal B}_2}{s}
\right.\nonumber\\&&\mbox{}\left.
+\left[\frac{\Box_3}{\Box_1\Box_2}(\frac12 {{\alpha_1}}^2)
-\frac{1}{\Box_1}({{\alpha_1}}^2)\right]\frac{{\cal A}_3}{s}
\right>_2
\nonumber\\&&\mbox{}
+\left<\frac{1}{\Box_1\Box_2}
\frac{{\rm e}^{s\Omega_3}-1}{s^2}
+\left[\frac{\Box_3}{\Box_1\Box_2}
\Big(3{\alpha_1}{\alpha_2}-2{\alpha_2})
\right.
\right.\nonumber\\& &\mbox{}\left.
+\frac{1}{\Box_2}
(-\frac{2}{3}+4{\alpha_2}+2{\alpha_1}-2{\alpha_2}^2-6{\alpha_1}{\alpha_2}
\Big)
\right]
\frac{{\rm e}^{s\Omega_3}}{s}
\nonumber\\& &\mbox{}
+\left[
\Big(-\frac{5}{36}+\frac{1}{2}{\alpha_1}
+\frac{2}{3}{\alpha_2}^2+\frac{2}{3}{\alpha_1}{\alpha_2}-{\alpha_2}^3
\right.\nonumber\\& &\mbox{}
-6{\alpha_1}{\alpha_2}^2+2{\alpha_1}{\alpha_2}^3+3{\alpha_1}^2{\alpha_2}^2
\Big)
\nonumber\\& &\mbox{}
+\frac{\Box_1}{\Box_2}
\Big(\frac{2}{3}{\alpha_1}{\alpha_2}
+\frac{1}{6}{\alpha_2}-{\alpha_2}^3
-3{\alpha_1}{\alpha_2}^2+2{\alpha_1}^2{\alpha_2}^2
\nonumber\\& &\mbox{}
-{\alpha_1}^2{\alpha_2}+2{\alpha_1}{\alpha_2}^3)
+\frac{\Box_3}{\Box_2}
\Big(-\frac{1}{6}{\alpha_2}
+{\alpha_2}^3-\frac{2}{3}{\alpha_1}{\alpha_2}
\nonumber\\& &\mbox{}
+4{\alpha_1}{\alpha_2}^2-4{\alpha_1}^2{\alpha_2}^2
-2{\alpha_1}{\alpha_2}^3+2{\alpha_1}^2{\alpha_2}\Big)
\nonumber\\& &\mbox{}
\left.\left.
+\frac{{\Box_3}^2}{\Box_1\Box_2}
({\alpha_1}^2{\alpha_2}^2-{\alpha_1}{\alpha_2}^2)
\right]
{\rm e}^{s\Omega_3}\right>_3,\label{G4a}
\\
G_5 &=&
\left< 2{1\over {\Box_1 \Box_2}}
\frac{{\rm e}^{s\Omega_3}-1}{s^2}
\right>_3
+\left<\left[{1\over {\Box_1}}(-{{{{\alpha_1}}^2}}) +
{{\Box_3}\over {\Box_1 \Box_2}}(\frac12{{{{\alpha_1}}^2}})\right]\right.
\frac{{\cal A}_3}{s}
\nonumber\\&&\mbox{}
+\left.\left({1\over {\Box_1}} -\frac32{{ \Box_3}\over { \Box_1 \Box_2}}
\right)
\frac{{\cal B}_3}{s}\right>_2,\label{G5a}
\\
G_6 &=&
\left< -2{{1}\over {\Box_1 \Box_2}}
\frac{{\rm e}^{s\Omega_3}}{s^2}+
\left[ {1\over {\Box_1}}(2 {\alpha_1} {\alpha_2}) + {1\over {\Box_2}}(2 {\alpha_1}
{\alpha_2})\right.\right.
\nonumber\\&&\mbox{}
 \left.\left.
+{{\Box_3}\over {\Box_1 \Box_2}}(-2 {\alpha_1} {\alpha_2}) \right]
\frac{{\rm e}^{s\Omega_3}}{s}\right>_3
+\left<{1\over {\Box_1 \Box_2}}({2 {\alpha_2}})
\frac{{\cal A}_3}{s^2}\right>_2,\label{G6a}
\\
G_7 &=&
-\left<
\frac{1}{\Box_2}(2{\alpha_2})
\rm{e}^{s\Omega_3}\right>_3+
\left<2\frac{1}{\Box_2}
\frac{{\cal U}_{13}}{s}\right>_2,\label{G7a}
\\
G_8 &=&
\left<\frac{1}{\Box_2}( 4{\alpha_2}^2) \rm{e}^{s\Omega_3}\right>_3
-\left<2\frac{1}{\Box_2}
\frac{{\cal U}_{13}}{s}\right>_2,\label{G8a}
\\
G_9 &=&
\left<
{1\over {\Box_1 \Box_2}}({8 {\alpha_1} {\alpha_2}})
\frac{{\rm e}^{s\Omega_3}}{s^2}
\right>_{3}
+\left<{1\over {\Box_1 \Box_2}}
(-{2 {{{\alpha_1}}^2}})
\frac{{\cal A}_{3}}{s^2}
\right.\nonumber\\&&\mbox{}
\left.
+2{1\over {\Box_1 \Box_2}}
\frac{{\cal B}_{3}}{s^2}\right>_2,\label{G9a}
\\
G_{10} &=&
\left<\frac{1}{\Box_1\Box_2}(4{\alpha_2}-12{\alpha_2}^2)
\frac{{\rm e}^{s\Omega_3}}{s^2}
+\left[\frac{1}{\Box_1}
({\alpha_1}{\alpha_2}^3-2{\alpha_1}{\alpha_2}^2)\right.\right.
\nonumber\\&&\mbox{}
+\frac{1}{\Box_2}
(\frac{4}{3}{\alpha_2}^2-4{\alpha_2}^3
+4{\alpha_2}^4+4{\alpha_1}{\alpha_2}^3-2{\alpha_1}{\alpha_2}^2)
\nonumber\\&&\mbox{}
+\left.\left.\frac{\Box_3}{\Box_1\Box_2}
(2{\alpha_1}{\alpha_2}^2-4{\alpha_1}{\alpha_2}^3)
\right]\frac{{\rm e}^{s\Omega_3}}{s}\right>_3
\nonumber\\&&\mbox{}
+\left<2\frac{1}{\Box_1\Box_2}
\frac{{\cal B}_{3}}{s^2}
+2\frac{1}{\Box_2}
\frac{{\cal V}_{13}}{s}
+\frac{1}{\Box_2}
\Big(\frac{1}{3}-2{\alpha_2}^2\Big)
\frac{{\cal U}_{13}}{s}
\right>_2,\label{G10a}
\\
G_{11} &=&
\left<
{1\over {\Box_1 \Box_2}}({4 {{{\alpha_1}}^2} {{{\alpha_2}}^2}})
\frac{{\rm e}^{s\Omega_3}}{s^2}\right>_3,\label{G11a}
\\
G_{12} &=&
\left<
2\frac{1}{\Box_1\Box_2}
\frac{{\rm e}^{s\Omega_3}}{s^2}
+\left[
\frac{\Box_3}{\Box_1\Box_2}
(2{\alpha_1}{\alpha_2})
-\frac{1}{\Box_2}
(2{\alpha_1}{\alpha_2})
\right.\right.\nonumber\\&&\mbox{}
\left.\left.
+\frac{1}{\Box_1}
(2{\alpha_1}-4{\alpha_1}^2-2{\alpha_1}{\alpha_2})
\right]
\frac{{\rm e}^{s\Omega_3}}{s^2}\right>_3
-\left<
\frac{1}{\Box_1\Box_2}
\frac{{\cal A}_3}{s^2}
\right>_2.\label{G12a}
\end{eqnarray}

The $\alpha$-polynomial representation for the  form factors
(\ref{G1a})--(\ref{G12a}) is not unique,
i.e., contributions of some form factors $G_i$, that  in fact 
identically vanish, can be present in the heat kernel. 
Due to this fact there is an extra quadratic structure (\ref{RRK12})
which will be  absent in the final answer for ${K}(s)$.
Thus, we need to proceed to the explicit representation
introduced in \cite{CPT4,CPT-book}. This has been done again in two ways,
the first is the treatment of (\ref{g1a})--(\ref{G12a})
with the $\alpha$~-polynomial reduction procedure derived in  \cite{CPT4,CPT-book}, see Appendix A.  The second is the use of expressions
(\ref{gf1})--(\ref{GF11}) and the data for the form factors
in the third curvature order of the heat kernel trace \cite{CPT4,CPT-book}.
Both methods result in the expression (\ref{K2}) for the heat kernel
where the contribution of the structure (\ref{RRK12}) vanishes.
The explicit form of the first order form factors 
is as follows \cite{BarGus-CQG92},
\begin{eqnarray}
g_1(\xi)&=&f(\xi), \label{g1}
\\
g_2({\xi_2})&=&\frac{1}{12}f(\xi)+\frac12\frac{f(\xi)-1}{\xi}.
\label{g2}
\end{eqnarray}
The structure and complexity of the next order
form factors are similar to those of the third order form factors
in the heat kernel trace. The full table of these form factors can be found in Appendix B. These tables, in form suitable for computer symbolic manipulations, can be obtained online from {\em Nuclear Physics B} or from the author.

The obvious check of the calculations above is
the functional trace operation (\ref{intro-8}).
We can use the $\alpha$~-representation of the form factors
for this consistency check.
All total derivative terms should be discarded, and all second order form factors reduced to first order ones by identities like
\begin{equation}
{\rm tr} \!
\int \! d \, x \,  g^{1/2}(x) F(-s\Box_1,-s\Box_2,-s\Box_3)
\Re_1 \Re_2=
\frac12 \, {\rm tr}\!
\int \! d x \,  g^{1/2}(x) f(-s\Box_2)
\Re_1 \Re_2 + {\rm O}[\Re^3].
\end{equation}
While first order terms collapse to a single local term $\hat{P}$,
the tables of the second order form factors reduce to
five form factors of the trace of the heat kernel found in \cite{CPT2}. 

Therefore, while performing the functional trace operation (\ref{intro-8})
after the generating function variation (\ref{varP}), we start with the higher order of the heat kernel trace and end up with the lower order one.
This shows that there exists a connection between two consecutive orders in the curvature of the heat kernel trace. Namely, each lower order in the curvatures is completely defined by the higher order. That is it,
\begin{equation}
{\cal K}_{n-1}={\rm tr} \int  dx
\frac{\delta}{\delta \hat{P}}\ {\cal K}_n,
\end{equation}
where
\begin{equation}
{\cal K}_n =
{\rm tr}\int  dx  g^{1/2}
\sum_i F_i(\nabla_1, \ldots , \nabla_n)
\Re_1 \ldots  \Re_n(i)
\end{equation}
is a given order in the curvatures.


\section{Short and large proper time behaviour of the heat kernel}
\label{shortK2}

Historically, a great deal of work on heat kernels has been devoted to
the short time expansion also known as the Schwinger-DeWitt (or HaMiDeW) series
\cite{DeWitt-book65,BarVilk-PRep85}:
\begin{equation}
\hat{K}(s|x,x)=\frac1{(4\pi s)^\omega}  g^{1/2}(x)
\sum_{n=0}^{\infty} s^n \hat{a}_n (x,x), \ \ \
s \rightarrow 0. \label{SDW1}
\end{equation}
The Schwinger-DeWitt coefficients  $\hat{a}_n (x,x)$
have been calculated explicitly for $n=0$ to $3$
in many works \cite{DeWitt-book65,BarVilk-PRep85,Gilk-JDG75}.
Several results for $n=4$ were obtained for special cases
\cite{WilkFujOsb-PRA81,AmstBerOcon-CQG89}, as well as for 
a general set of fields \cite{Avram-NPB91,VdeVen-NPB98}.
 \cite{VdeVen-NPB98} gives even the fifth coefficient, but
its expressions are written in the index free form, 
therefore, it would require a special study to transform it 
into a representation suitable for our comparative analysis.
All $\hat{a}_n (x,x)$ coefficients are local functions of the
curvatures  (\ref{intro-6}).
The short time expansion serves as a limiting case and as
a consistency check for the results presented in the previous section.

To find the short time asymptotic expansion (\ref{SDW1}),
we need to know short time expansions of
the form factors in $\hat{K}(s|x,x)$.
Generally, this procedure is very much similar to the one performed 
for the trace of the heat kernel in \cite{CPT4,BGVZ-JMP94-asymp}.
It is a simple exercise  to expand the  basic form factors (\ref{secondff})-(\ref{alpha-12}).
By inserting these expansions into  the table of form factors, we obtain
the Schwinger-DeWitt series. Since $\hat{a}_n (x,x)$ are local functions of the background fields,
the nonlocal denominators $1/\Delta$ cancel in
the short time expansions of form factors, as expected.
Still, the tree nonlocal terms $1/\Box$ are present
due to the Riemann tensor expansion in terms of the Ricci tensor \cite{CPT1}.
We omit here these nonlocal expressions for the Schwinger-DeWitt coefficients.

Technically, the tree nonlocal expressions appear due to the Bianchi identity \cite{CPT2}
\begin{eqnarray}&& 
R^{\alpha\beta\mu\nu}=
\frac1{\Box}\Big(\nabla^\mu \nabla^\alpha R^{\nu\beta}
-\nabla^\nu \nabla^\alpha R^{\mu\beta}
-\nabla^\mu \nabla^\beta R^{\nu\alpha}
+\nabla^\nu \nabla^\beta R^{\mu\alpha}\Big)
+{\rm O}[R^2].  
\label{Riemann}
\end{eqnarray}
Since we are working in the second order in the curvatures, we also need the nonlocal expansion for the square of the Riemann tensor:
\begin{eqnarray}
&& R_{\mu\nu\alpha\beta} R^{\mu\nu\alpha\beta}=
4 R_{\mu\nu\alpha\beta}
\nabla_{\mu} \nabla_{\alpha} \frac1{\Box} R_{\nu\beta}
+{\rm O}[R^3]
\nonumber\\&&\mbox{} =\frac1{\Box_1\Box_2}
\nabla^{\nu} \nabla^{\beta} R^{\mu\alpha}_1
\nabla_{\mu} \nabla_{\alpha} R_{2\nu\beta}
+ \frac1{\Box_1\Box_2} (\Box_1 + \Box_2 - \Box_3)
\nabla^{\beta} R^{\mu\alpha}_1
\nabla_{\alpha} R_{2\mu\beta}
\nonumber\\&&\mbox{}
+ \frac14 \frac1{\Box_1\Box_2}
({\Box_1}^2 + {\Box_2}^2 + {\Box_3}^2  + 2 \Box_1 \Box_2 -
2 \Box_1 \Box_3 -  2 \Box_2 \Box_3 )
 R^{\mu\nu}_1 R^{\mu\nu}_2
\nonumber\\&&\mbox{}
+{\rm O}[R^3].  \label{Riem2}
\end{eqnarray}

The true local representation of the derived short time expansion for the nonlocal heat kernel can be restored, if we find an irreducible basis of  quadratic tensor invariants containing the Riemann tensor, for each of the coefficients. It is easy to do in the second order, as the basis for $a_2(x,x)$
consists of just one structure,
\begin{equation}
R_{\mu\nu\alpha\beta}R^{\mu\nu\alpha\beta}.  \label{loc2}
\end{equation}
Since there are several derivatives eliminating the Riemann tensor with help of the Bianchi identity
in the third and fourth coefficients, the basis for $a_3(x,x)$ may contain only
\begin{equation}
\Box (R_{\mu\nu\alpha\beta}R^{\mu\nu\alpha\beta}),
\ \ \ \
R_{\mu\nu\alpha\beta} \nabla^{\mu} \nabla^{\alpha} R^{\nu\beta}, \label{loc3}
\end{equation}
and for $a_4(x,x)$,
\begin{equation}
\Box^2 (R_{\mu\nu\alpha\beta}R^{\mu\nu\alpha\beta}),
\ \ \ \
\Box( R_{\mu\nu\alpha\beta} \nabla^{\mu} \nabla^{\alpha} R^{\nu\beta}),
\ \ \ \
\Box( R_{\mu\nu\alpha\beta} \Box \nabla^{\mu} \nabla^{\alpha} R^{\nu\beta}).
\label{loc4}
\end{equation}
We construct now the local $a_n(x,x)$ with help of
the invariants (\ref{loc2})-(\ref{loc4}) and other acceptable tensor invariants
(\ref{RRK1})--(\ref{RRK11}), mixed with a proper number of the Laplacians $\Box$.
These expressions are substituted into constructed local $a_n(x,x)$
with unknown numerical coefficients.
Equating the resulting quantities to the short time expansion in its nonlocal representation
we arrive at a unique solution for the numerical coefficients to be found.

The final result is,
\begin{eqnarray}
\hat{a}_2(x,x)&=& \frac16 \Box\hat{P}+\frac1{180}\Box R \hat{1}
+\frac12 \hat{P}\hat{P}+
\frac1{12}{\cal R}_{\mu\nu}{\cal R}^{\mu\nu}
\nonumber\\&&\mbox{}
+\left[\frac1{180} R_{\alpha\beta\mu\nu}R^{\alpha\beta\mu\nu}
-\frac1{180} R_{\mu\nu}R^{\mu\nu}\right]\hat{1}
+{\rm O}[\Re^3],  \label{a2loc}
\\
\hat{a}_3(x,x)&=&
\frac{1}{60}{\Box}^2\hat{P}
+\frac{1}{1260}{\Box}^2 R\hat{1}
\nonumber\\&&\mbox{}
+\frac{1}{24} \Box (\hat{P} \hat{P})
+\frac{1}{24} \Box \hat{P} \hat{P}
+\frac{1}{24} \hat{P} \Box \hat{P}
\nonumber\\&&\mbox{}
+\frac{1}{90}\Box ( \hat{\cal R}^{\mu\nu} \hat{\cal R}_{\mu \nu})
+\frac{1}{180} \Box \hat{\cal R}^{\mu\nu} \hat{\cal R}_{\mu \nu}
+\frac{1}{180}  \hat{\cal R}^{\mu\nu}\Box \hat{\cal R}_{\mu \nu}
\nonumber\\&&\mbox{}
+ \frac{1}{360}\Box(\hat{P} R)
-\frac{1}{360}\Box \hat{P} R
+\frac{1}{360}\hat{P}\Box R
\nonumber\\&&\mbox{}
+ \frac{1}{180}
\nabla_{\mu}\hat{\cal R}^{\mu\nu} \nabla^{\alpha}\hat{\cal R}_{\alpha\nu}
-\frac{1}{60}
\Big[\nabla_{\alpha}\hat{P},\nabla_{\beta}\hat{\cal R}^{\beta\alpha}\Big]
+\frac{ 1}{90}
 \nabla_\mu \nabla_\nu \hat{P} R^{\mu\nu}
\nonumber\\&&\mbox{}
+\left[
\frac{1}{1120}\Box (
R^{\mu\nu\alpha\beta} R_{\mu\nu\alpha\beta})
+ \frac{1}{420}
R^{\mu\nu\alpha\beta} \nabla_{\mu}\nabla_{\alpha}
R_{\nu\beta}
-\frac{1}{840}
R^{\mu\nu}\Box R_{\mu\nu}\right.
\nonumber\\&&\mbox{}
-\frac{1}{5040}
\Box(R^{\mu\nu} R_{\mu\nu})
-\frac{1}{1260}
\nabla^{\mu}R^{\nu\alpha}
\nabla_{\alpha}R_{\mu\nu}
+ \frac{1}{1890}
\nabla_\mu \nabla_\nu R R^{\mu\nu}
\nonumber\\&&\mbox{}\left.
-\frac{1}{7560} R\Box R
+\frac{1}{15120}\Box( R R)\right]
+{\rm O}[\Re^3],  \label{a3loc}
\\
\hat{a}_4(x,x)&=&
\frac{1}{840}{\Box}^3\hat{P}+\frac{1}{15120}{\Box}^3 R\hat{1}
\nonumber\\&&\mbox{}
+\frac{1}{360} {\Box}( \hat{P}\Box \hat{P})
+\frac{1}{360}\Box \hat{P}\Box \hat{P}
+\frac{1}{360}\Box(\Box\hat{P} \hat{P})
\nonumber\\&&\mbox{}
+\frac{1}{360}{\Box}^2(\hat{P} \hat{P})
+\frac{1}{360}{\Box}^2\hat{P} \hat{P}
+\frac{1}{360} \hat{P}{\Box}^2 \hat{P}
\nonumber\\&&\mbox{}
+\frac{1}{3360}{\Box}^2
\hat{\cal R}^{\mu\nu} \hat{\cal R}_{\mu \nu}
+\frac{1}{3360}
\hat{\cal R}^{\mu\nu}{\Box}^2 \hat{\cal R}_{\mu \nu}
+\frac{1}{2520}
{\Box}\hat{\cal R}^{\mu\nu}{\Box} \hat{\cal R}_{\mu \nu}
\nonumber\\&&\mbox{}
+\frac{1}{1680}
{\Box}({\Box}\hat{\cal R}^{\mu\nu} \hat{\cal R}_{\mu \nu})
+\frac{1}{1120}
{\Box}^2(\hat{\cal R}^{\mu\nu} \hat{\cal R}_{\mu \nu})
+\frac{1 }{1680}
{\Box}(\hat{\cal R}^{\mu\nu} {\Box}\hat{\cal R}_{\mu \nu})
\nonumber\\&&\mbox{}
+\frac{1}{15120}{\Box}({\Box}\hat{P} R)
-\frac{1}{3024} {\Box}^2\hat{P} R
+\frac{1}{3780}\hat{P}{\Box}^2 R
\nonumber\\&&\mbox{}
+\frac{1}{3780}{\Box}^2( \hat{P} R)
+\frac{1}{15120}{\Box}\hat{P}{\Box} R
+\frac{1}{3780}{\Box}(\hat{P}{\Box} R)
\nonumber\\&&\mbox{}
+\frac{1}{2520}{\Box}(
\nabla_{\mu}\hat{\cal R}^{\mu\nu} \nabla^{\alpha}\hat{\cal R}_{\alpha\nu})
+\frac{1}{2520}
\nabla_{\mu}\hat{\cal R}^{\mu\nu}\Box \nabla^{\alpha}\hat{\cal R}_{\alpha\nu}
+\frac{1}{2520}\Box
\nabla_{\mu}\hat{\cal R}^{\mu\nu} \nabla^{\alpha}\hat{\cal R}_{\alpha\nu}
\nonumber\\&&\mbox{}
-\frac{1}{630}
\big[\Box\nabla_{\alpha}\hat{P},\nabla_{\beta}\hat{\cal R}^{\beta\alpha}\big]
-\frac{1}{630} \Box
\big[\nabla_{\alpha}\hat{P},\nabla_{\beta}\hat{\cal R}^{\beta\alpha}\big]
-\frac{1}{1260}
\big[\nabla_{\alpha}\hat{P},\Box\nabla_{\beta}\hat{\cal R}^{\beta\alpha}\big]
\nonumber\\&&\mbox{}
+\frac{1}{840} \Box \nabla_\mu \nabla_\nu \hat{P} R^{\mu\nu}
+\frac{1}{2520} \nabla_\mu \nabla_\nu \hat{P} \Box R^{\mu\nu}
+\frac{1}{840} \Box( \nabla_\mu \nabla_\nu \hat{P} R^{\mu\nu})
\nonumber\\&&\mbox{}
+\left[\frac{1}{12600}
{\Box}^2
(R^{\mu\nu\alpha\beta} R_{\mu\nu\alpha\beta})
+\frac{1}{3150}
{\Box}
( R^{\mu\nu\alpha\beta}\nabla_\mu \nabla_\alpha R_{\nu\beta})
+\frac{1}{9450}
R^{\mu\nu\alpha\beta}
\Box
\nabla_\mu \nabla_\alpha R_{\nu\beta}
\right.\nonumber\\&&\mbox{}
+\frac{1}{12600}
\nabla_\alpha \nabla_\beta R_{\mu\nu}
\nabla^\mu \nabla^\nu R^{\alpha\beta}
-\frac{1}{6300}\Box
(\nabla_\alpha R_{\mu\nu} \nabla^{\mu}R^{\nu\alpha})
+\frac{1}{18900}
\nabla_\alpha R_{\mu\nu}
\Box
\nabla^{\mu}R^{\nu\alpha}
\nonumber\\&&\mbox{}
+\frac{1}{15120}
\Box
(\nabla_\mu \nabla_\nu R R^{\mu\nu})
+\frac{1}{15120}
\Box\nabla_\mu \nabla_\nu R R^{\mu\nu}
+\frac{1}{75600}
\nabla_\mu \nabla_\nu R \Box R^{\mu\nu}
\nonumber\\&&\mbox{}
+\frac{1}{50400}
{\Box}^2
(R^{\mu\nu} R_{\mu\nu})
-\frac{11}{75600}
{\Box}
(R^{\mu\nu} \Box R_{\mu\nu})
-\frac{1}{37800}
R^{\mu\nu}{\Box}^2 R_{\mu\nu}
\nonumber\\&&\mbox{}
-\frac{1}{75600}
{\Box}
R^{\mu\nu}{\Box} R_{\mu\nu}
 -\frac{1}{56700}  {\Box}^2 R R
+\frac{1}{453600}  {\Box}( {\Box} R R )
\nonumber\\&&\mbox{} \left.
-\frac{1}{453600}  {\Box}R  {\Box}R
+\frac{1}{129600}  {\Box}^2(R R) \right]\hat{1}
+{\rm O}[\Re^3].   \label{a4loc}
\end{eqnarray}
Two terms in Eq.~(\ref{a4loc}), $R^{\mu\nu}{\Box}^2 R_{\mu\nu}$
and ${\Box} R^{\mu\nu}{\Box} R_{\mu\nu}$,
have their numerical coefficients corrected from \cite{BarGus-Winn94}.
Although $a_2(x,x)$ was computed with ${\rm O}[\Re^3]$ accuracy,
its expression above is, of course, exact.

Now we compare our results with the coefficients available
in the literature.  The main concern is the fourth coefficient $a_4(x,x)$, as
expressions (\ref{a2loc}), (\ref{a3loc}) for $a_2(x,x)$
and $a_3(x,x)$ coincide with the results obtained by other methods
\cite{DeWitt-book65,Avram-NPB91,Gilk-JDG75},
after taking into account differences in definitions and curvature conventions.
In  \cite{BranGilkOrst-PAMS90}, the first order curvature terms are presented
for all Schwinger-DeWitt coefficients.
This result is fully consistent  with similar terms in coefficients (\ref{a2loc})--(\ref{a4loc}).

The only general result for the fourth Schwinger-DeWitt coefficient $a_4(x,x)$ belongs to
I. Avramidi \cite{Avram-NPB91}. Unfortunately, his result is not expressed in terms of an irreducible basis of tensor invariants.
The tedious work of the reduction to a tensor invariant basis, 
that we choose to be the basis above, was performed. It was done partially with help of 
tensor manipulation software packages, {\em MathTensor} \cite{MathTensor} and {\em Ricci} \cite{Ricci}.
This reduction is even more difficult than a similar procedure for
$\int \! d x \, {\rm tr}a_4(x,x)$ performed in \cite{BGVZ-JMP94-asymp} because of the presence of total derivatives.
For example, the result of \cite{Avram-NPB91} contains a term which represents six covariant derivatives 
acting on the Ricci tensor, totally symmetrized over indeces . It was reduced to the term  $\Box^3 R$, while keeping all curvature terms arising from the derivatives commuting. The use of computers for such calculations is unavoidable,
and the ability of {\em Ricci} to perform index free manipulations is invaluable.
The result derived this way from  \cite{Avram-NPB91} is in a full agreement with (\ref{a4loc}).
It should be noted, that we were only concerned with reproducing the squared curvatures terms, and the complete fourth Schwinger-DeWitt coefficient in a form compatible with the lower order ones remains to be derived from either \cite{Avram-NPB91} or \cite{VdeVen-NPB98}.

The work \cite{AmstBerOcon-CQG89} gives only pure gravitational terms, thus, only
terms of (\ref{a4loc}) in the square brackets can be compared with it.
Our expression (\ref{a4loc}) disagrees with \cite{AmstBerOcon-CQG89} in several numerical coefficients.
Taking into account, that results of  \cite{AmstBerOcon-CQG89}
fail to reproduce  correctly even the lowest order terms of the trace of the fourth Schwinger-DeWitt 
coefficient, this work should be considered erroneous.

We also mention that it might be possible to obtain the Schwinger-Dewitt
coefficients above from the short time expansion
of the trace of the heat kernel found in \cite{BGVZ-JMP94-asymp}
by applying the variational principle (\ref{varP}).
However, in order to derive the fourth coefficient $a_4(x,x)$,
that expansion should have been derived up to the fifth order in the proper time.
This fact follows directly from  the Schwinger-DeWitt expansion (\ref{SDW1})
and (\ref{varP}),
\begin{equation}
\hat{a}_{n-1}(x,x)=g^{-1/2}
\frac{\delta}{\delta \hat{P}}\ 
\int dx \ g^{1/2}{\rm tr} \ \hat{a}_n(x,x),
\end{equation}
and was used in the gauge field theory \cite{Ball-PRep89}.

It is even easier to derive the large (sometimes called late) proper time asymptotic of the heat kernel. The large time expansions of the basic form factors are known \cite{BGVZ-JMP94-asymp},
\begin{equation}
f(-s\Box) = -\frac1s\frac2{\Box}
+{\rm O}\left(\frac1{s^2}\right),
\hspace{7mm} s\rightarrow\infty,    \label{large-1}
\end{equation}
\begin{equation}
F(-s\Box_1,-s\Box_2,-s\Box_3) =
\frac1{s^2}\Big(\frac1{\Box_1\Box_2}+\frac1{\Box_1\Box_3}+
\frac1{\Box_2\Box_3}\Big)
+{\rm O}\left(\frac1{s^3}\right),
\hspace{7mm} s\rightarrow\infty.  \label{large-2}
\end{equation}
Their substitution to the expressions for the heat kernel's form factors leads to the following equation,
\begin{eqnarray}
\hat{K}(s|x,x)&=&\frac{1}{(4\pi s)^{\omega}}
g^{1/2}
\Big\{\hat{1}
- 2 \frac{1}{\Box}\hat{P} + \frac{1}{3} \frac{1}{\Box}R\hat{1}
\nonumber\\&&
+ 2 \left( \frac{1}{\Box} \hat{P} \right) \left( \frac{1}{\Box} \hat{P} \right)
+  \frac{1}{\Box} \left( \hat{P}  \frac{1}{\Box} \hat{P} \right)
\nonumber\\&&
+ \frac{1}{36} \left( \frac{1}{\Box} R \right) \left( \frac{1}{\Box} R \right) \hat{1}
+ \frac{1}{18} \frac{1}{\Box}  \left( R \frac{1}{\Box} R \right) \hat{1}
\nonumber\\&&
- \frac{1}{3} \left( \frac{1}{\Box} \hat{P} \right) \left( \frac{1}{\Box} R \right)
- \frac{1}{3} \frac{1}{\Box} \left( \hat{P} \frac{1}{\Box} R \right)
- \frac{1}{3} \frac{1}{\Box} \left( R \frac{1}{\Box} \hat{P} \right)
\nonumber\\&&
+ 2 \frac{1}{\Box} \left(
 \nabla_{\alpha} \frac{1}{\Box} \hat{P} \
 \nabla_{\beta} \frac{1}{\Box} \hat{\cal R}^{\beta\alpha}\right)
 - 2 \frac{1}{\Box} \left(
  \nabla_{\beta} \frac{1}{\Box} \hat{\cal R}^{\beta\alpha}\
 \nabla_{\alpha} \frac{1}{\Box} \hat{P} 
 \right)
 \nonumber\\&&
 - 2  \frac{1}{\Box} \left(
\nabla_{\mu}\frac{1}{\Box}\hat{\cal R}^{\mu\nu} \
\nabla^{\alpha}\frac{1}{\Box}\hat{\cal R}_{\alpha\nu}
\right)
+\rm{O}[\Re^3]\Big\}.         \label{largetime}
\end{eqnarray}
This formula is in agreement with the result of variation of the large time asymptotic of the trace of the heat kernel found in \cite{BGVZ-JMP94-asymp}.


\section{Green function for the massless scalar field in two dimensions} 
\label{Weyl}

As a test of our results for the heat kernel,
in this section we derive the Green function 
of the scalar field model in two dimensions.
The curvature expansion for the one-loop effective action (\ref{intro-7})
in two dimensions does not exist 
due to infrared divergences appearing at every order in the curvature
due to Eq.~(\ref{trklarges}).
The exception is the two-dimensional effective action for massless fields \cite{Pol-PLB81},
which does exist and admits an exact form,
\begin{equation}
W= \frac{1}{96 \pi}\int dx g^{1/2}R\frac{1}{\Box} R. \label{Wd2}
\end{equation}
How this effective action arises from the covariant perturbation theory was shown in \cite{CPT2,BGVZ-NPB95}.

Naturally, we expect this behaviour to hold also  for the corresponding Green functions 
(\ref{propagator})-(\ref{Schwinger}).
As shown in \cite{Pol-PLB81}, we can start by deriving 
the exact two-dimensional Green function  from the heat kernel in a flat space-time (\ref{K0}).
Then we performs the Weyl transformation \cite{FradVilk-PLB78}, which in two dimensions consists of only the local rescaling of the metric,
\begin{equation}
\tilde{g}^{\mu\nu}(x) = {\rm e}^{\Sigma (x)}g^{\mu\nu}(x), \label{Weylg}
\end{equation}
where $\Sigma(x)$ is a smooth function vanishing at space-time infinity. 
The outcome would simply be,  
\begin{equation}
G(x,x)=-\frac1{4\pi}g^{1/2}(x)\Sigma(x). \label{Gd2}
\end{equation}
This expression admits a nonlocal closed form,
upon inserting the conformal factor $\Sigma$
expressed in terms of the covariant metric
\begin{equation}
\Sigma(g)=\frac{1}{\Box} R. \label{sigmag}
\end{equation}

In order to arrive at this result, we proceed first by formally defining the regularized 
Green function via the heat kernel as,
\begin{equation}
G^{\rm reg}=  \int_0^{\infty} \! d\, s \Big[
{K}(s)- {K}(s)|_{\Re=0}\Big],  \label{gr}
\end{equation}
and then computing with the use of the explicit form of the heat kernel from  the section \ref{explicitK2}.

There is only one curvature in two dimensions due to the properties,
\begin{equation}
{\rm tr} \hat{1}=1,\hspace{7mm}  \hat{\cal R}_{\mu\nu}=0, \hspace{7mm} R_{\mu\nu}=\frac12 g_{\mu\nu}R,
\hspace{7mm} \hat{P}=\frac16 R\hat{1}. \label{twod-2}
\end{equation}
Therefore, the heat kernel (\ref{K2}) reduces to:
\begin{eqnarray}
K(s)&=&\frac{g^{1/2}}{4 \pi s}
          \left\{1+ s \sum_{i=1}^2 \tilde{c}_i g_i(-s{\Box})R\right.
\nonumber\\&&\left. \mbox{}
+ s^2 \sum_{i=1}^{11}  \tilde{C}_i\ G_i(-s\Box_1,-s\Box_2,-s\Box_3)
R_1 R_2 +{\rm O}[R^3]\right\},
\end{eqnarray}
where the coefficients $\tilde{c}_i$ and $\tilde{C}_i$ are
\begin{eqnarray}&&
\tilde{c}_1=\frac16, \ \ \ \tilde{c}_2=1,
\nonumber\\&&
\tilde{C}_1=\frac{1}{36}, \ \ \tilde{C}_3=\frac16, \ \ \tilde{C}_4=1, \ \
\tilde{C}_5=\frac12, \ \ \tilde{C}_8=\frac{s}{12}\Box_1, \ \
\tilde{C}_9=\frac{s}{8}(\Box_3-\Box_2-\Box_1), 
\nonumber\\&&
\tilde{C}_{10}=\frac{s}{2}\Box_1, \ \
\tilde{C}_{11}=\frac{s^2}{4}\Box_1\Box_2,\ \
\tilde{C}_2=\tilde{C}_6=\tilde{C}_7=0.
\end{eqnarray}

The final expression looks like
\begin{eqnarray}
K(s)- K(s)|_{R=0}&=&\frac{g^{1/2}}{4 \pi}
          \left\{\left[
\frac{1}{4}f(-s{\Box})
-\frac12 \frac{f(-s{\Box})-1}{s{\Box}}\right]R
\right.\nonumber\\&&\mbox{}
+ \left[sF(-s\Box_1,-s\Box_2,-s\Box_3)
\frac{\Box_1 \Box_2 {\Box_3}^2}{D^2} \right.
\nonumber\\&&\mbox{}
+f(-s\Box_1) \frac{\Box_1}{4 \Box_2 {{D}}^2} 
({\Box_1}^3-3 \Box_2 {\Box_1}^2-3 \Box_3 {\Box_1}^2
+3 \Box_1 {\Box_2}^2
\nonumber\\&&\mbox{}
-2 \Box_1 \Box_2 \Box_3
+3 \Box_1 {\Box_3}^2+5 \Box_2 {\Box_3}^2
-{\Box_2}^3+5 \Box_3 {\Box_2}^2-{\Box_3}^3)
\nonumber\\&&\mbox{}
- f(-s\Box_3) \frac{1}{4 {{D}}^2 \Box_2}
(-\Box_1 {\Box_3}^3+3 {\Box_1}^2 {\Box_3}^2
-3 \Box_3 {\Box_1}^3+{\Box_1}^4-\Box_2 {\Box_3}^3
\nonumber\\&&\mbox{}
+9 \Box_1 \Box_2 {\Box_3}^2
-4 {\Box_1}^2 \Box_2 \Box_3-3 \Box_2 {\Box_1}^3
+7 \Box_1 \Box_3 {\Box_2}^2+2 {\Box_1}^2 {\Box_2}^2)
\nonumber\\&&\mbox{}
+ \left(\frac{f(-s{\Box_1})-1}{s{\Box_1}}\right)
\frac{\Box_1}{2 \Box_2 {D}}(\Box_3+\Box_2-\Box_1)
\nonumber\\&&\mbox{}
+\left(\frac{f(-s{\Box_3})-1}{s{\Box_3}}\right)
\frac{1}{2 \Box_2 {D}}
(-\Box_1 \Box_3+{\Box_1}^2-3 \Box_2 \Box_3-\Box_1  \Box_2)
\nonumber\\&&\mbox{}
-\frac{1}{({\Box_1}-{\Box_3})}
\Big(f(-s{\Box_1})-f(-s{\Box_3})\Big)
\frac{\Box_1}{4 \Box_2}
\nonumber\\&&\left.\mbox{}
+\frac{1}{({\Box_1}-{\Box_3})}
\left(\frac{f(-s{\Box_1})-1}{s{\Box_1}}
-\frac{f(-s{\Box_3})-1}{s{\Box_3}}\right)
\frac{\Box_1}{2 \Box_2}
\right] R_1 R_2
\nonumber\\&&\left.\mbox{}
 + {\rm O}[R^3] \right\}. \label{K2-2D}
\end{eqnarray}
A straightforward check with help of the large time asymptotics
(\ref{large-1})-(\ref{large-2})
shows that the expression (\ref{K2-2D}) behaves like $s^{-2}$
at $s \rightarrow \infty$, which provides the convergence of 
the proper time integral (\ref{gr}).
To perform the proper time integration 
we resort to the technique of \cite{CPT-book,BGVZ-JMP94-asymp},
namely, we express $K(s)$ as a total derivative over  the proper time:
\begin{equation}
K(s) - K(s)|_{R=0}=\frac{g^{1/2}}{4 \pi}
          \frac{\rm d}{{\rm d} s}\Big(
 n(s,\Box) + 
m^{\rm sym}(s,\Box_1, \Box_2, \Box_3) \Big) + {\rm O}[R^3],
\label{totald}
\end{equation}
where
\begin{equation}
n(s,\Box) = 
\frac{1}{\Box}f(-s \Box)R,
\end{equation}
and $m^{\rm sym}$
is the following function symmetrized over indices 1 and 2:
\begin{eqnarray}
m(s,\Box_1, \Box_2, \Box_3) &=&
\left[- s \frac{\Box_3}{\Box_2 D}  F(-s\Box_1,-s\Box_2,-s\Box_3)
\right.\nonumber\\&&\mbox{}
-f(-s\Box_1)\frac{(\Box_3+\Box_2-\Box_1)}{\Box_2 D}
\nonumber\\&&\mbox{}
-f(-s\Box_3)\frac{(- \Box_1 \Box_3+{\Box_1}^2 
-3 \Box_2 \Box_3 -\Box_1 \Box_2)}{\Box_2 \Box_3D}
\nonumber\\&&\left. \mbox{}
+ \frac{1}{({\Box_1}-{\Box_3})}
\frac{\Box_1}{\Box_2}
\left(\frac{f(-s{\Box_3})}{{\Box_3}}
-\frac{f(-s{\Box_1})}{{\Box_1}}\right)
\right]R_1 R_2.
\label{mfactor}
\end{eqnarray}
After Substituting the expansions (\ref{large-1})--(\ref{large-2}) 
into (\ref{totald}) and doing the proper time integration, 
we observe that the whole expression vanishes at large $s$, and the resulting equation is
\begin{equation}
G^{\rm reg}(x,x)=-\frac{g^{1/2}}{4 \pi} 
\Big( n(0,\Box) + m^{\rm sym}(0,\Box_1, \Box_2, \Box_3)
\Big) + {\rm O}[R^3]. 
\end{equation}
As is seen from (\ref{mfactor}), only the linear in curvature term survives, so,
the Green function we seek is
\begin{equation}
G^{\rm reg}(x,x)= -\frac{g^{1/2}}{4 \pi} \frac{1}{\Box} R + {\rm O}[R^3]. 
\label{1/F}
\end{equation}

Although we have only demonstrated that the second order 
in the curvature expansion is exactly zero,
it is clear from (\ref{Gd2})--(\ref{sigmag}) 
that the answer (\ref{1/F}) is, in fact, exact,
in the same fashion as the one-loop effective action 
(\ref{Wd2}) is exact in two dimensions \cite{CPT2,BGVZ-NPB95}.
We note, however, that the expression (\ref{1/F}) 
cannot be directly obtained from the corresponding effective action (\ref{Wd2})
by the generating function method \cite{BarVilk-QFTQS87} because there is
no potential term $\hat{P}$ with respect 
to which we can take a variational derivative.

It is appropriate here to remind one how to understand and work with the results of this paper or any paper on the covariant perturbation theory,
\begin{equation}
    \frac1{\Box}R(x)\equiv
    \int dy\,
    G(x,y)\,R(y),
\end{equation}
where the Green function $G(x,y)$ of $\Box$ is defined as,
\begin{equation}
\Box_x G(x,y)= \delta(x,y), \ \ \
G(x,y)\,\big|_{\,|x|\to\infty}=0.
\end{equation}


\section{Summary}

Let us summarize this study. In this work we have implemented the covariant perturbation theory for the coincidence limit  of the heat kernel. It was computed
up to the second order in the curvatures by the generating function
method from the known heat kernel trace, and also 
directly by the covariant perturbation theory algorithms.
The basis of the second order nonlocal tensor invariants was obtained.
The nonlocal form factors are given in the two integral representations.
By taking the functional trace of the heat kernel, 
the lower, known order of the heat kernel trace is reproduced.
We have compared the short proper time expansion of
our results with the Schwinger-DeWitt coefficients found in the literature.
The work \cite{Avram-NPB91} is confirmed,
while the other \cite{AmstBerOcon-CQG89} is proved to be erroneuos.
The diagonal Green function for the massless scalar field model in two dimensions was derived and shown to produce the closed form nonlocal expression.
Due to these tests, the results of this paper can be deemed,
via the variational principle, another verification  of the third order of the trace of the heat kernel derived in 
\cite{CPT-book,BGVZ-JMP94-bas,BGVZ-JMP94-asymp}.

The validity of the covariant perturbation theory obeys the condition
\cite{CPT2},
\begin{equation}
  \nabla \nabla {\Re} >> \Re^2.
\end{equation}
This means is the results are applicable to problems with rapidly oscillating
background fields of small magnitudes, i.e., the high frequency limit.
The opposite case of slowly fluctuating fields of large magnitudes, 
which corresponds to the effective potential, 
also presents a significant interest and was studied in
\cite{Avram-JMP95}.

The obtained results can be directly applied to the computation of the Green function
in four dimensions. To derive it, one needs  to make an additional step,
the integration over the proper time (\ref{propagator}).
It is proved that only the first order in curvatures is ultraviolet divergent 
\cite{DeWitt-book65,BarVilk-PRep85},
which in the dimensional regularization \cite{tHooft-NPB73} is
\begin{equation}
\hat{G}(x,x)^{\rm div} =
\frac{1}{2-\omega}\frac1{16 \pi^2}\,  g^{1/2}\hat{P},
\hspace{15mm} \omega \rightarrow 2,
\end{equation}
while the second and higher orders are finite and nonlocal. 
As was mentioned, infrared divergences of massless field 
theories do not appear at all due to the nice behavior 
of the heat kernel at large proper times (cf., Eq.~(\ref{trklarges})),
\begin{equation}
K(s) \propto s^{-\omega},\ \ \ s \rightarrow \infty,  \ \ \ \Re \neq 0.
\label{larges}
\end{equation}
which holds for any curvature order excluding the zeroth.

The found nonlocal contributions of the Green functions
are responsible for a variety of physical effects \cite{Vilk-Gospel}.
All physical information is contained in the nonlocal form factors.
However, one needs to transform these form factors to
the massive Green functions \cite{CPT1,CPT2,CPT3,CPT4}.
The first order form factors turn, upon the proper time integration, to
\begin{equation}
- \ln \left(\frac{-\Box}{\ \mu^2}  \right) = \int_{0}^{\infty} d m^2 \left(\frac{1}{m^2-\Box}
-\frac{1}{m^2+\mu^2} \right),
\end{equation}
where $\mu^2$ is the parameter of ultraviolet regularization.
For the second order form factors, the following construction can
be derived \cite{CPT3},
\begin{equation}
\int_{0}^{\infty} \frac{d {m_1}^2 d {m_2}^2 d{m_3}^2
		\rho (m_1, m_2, m_3)}
{({m_1}^2-\Box)   ({m_2}^2-\Box)     ({m_3}^2-\Box)}.
\end{equation}
These spectral forms encode the information about
the filed model into a set of spectral densities $\rho({m_1}^2,{m_2}^2,{m_3}^2)$ 
\cite{Vilk-CQG92}.
Another step towards physical applications is going from the Euclidean signature metric 
to the Minkowskian one. The unique Green function of the Euclidean space gets replaced by one of many. Which one is being defined by the physical problem under consideration via the integration contours \cite{DeWitt-book65}.
If one is solving a Cauchy problem with data in the remote past,
for expectation values of fields in the standard {\em in}-vacuum,
this procedure boils down to using the retarded massive Green functions 
in the  effective equations \cite{CPT1,FrolVilk-PLB81,GAV-lectures07}.

The very first application of the nonlocal effective action was quantum electrodynamics where the correct derivation of the electron magnetic moment, free of infrared divergences, was done \cite{OstrVilk-JMP88}.
General results of the covariant perturbation theory are quite universal and applicable to a wide class of physically interesting quantum field theories that are specified by the differential operator  $\hat{F}(\nabla)$ of the form (\ref{intro-2}).
Starting from the model's Lagrangian one can write down $\hat{F}(\nabla)$ 
and explicit forms for the curvatures  $\hat{P}$ and $\hat{\cal R}_{\mu\nu}$.
The prescriptions on reducing various field models to the standard form (\ref{intro-2}) can be found in \cite{BarVilk-PRep85}.

For the example of quantum electrodynamics coupled to gravity, 
the whole structure of the nonlocal effective action was already worked out by G. Shore \cite{GShore-NPB02}. 
It seems instructive to show here explicit forms for the CPT curvatures of this model, omitting the model's description and explanations that can be found in \cite{GShore-NPB02}. The potential term is 
$\hat{P} = i e \sigma^{\mu\nu} F_{\mu\nu} - \frac{1}{12} R \hat{1}$, and the commutator curvature takes the form of
$\hat{\cal{R}}_{\mu\nu}= i e F_{\mu\nu} \hat{1} + \frac12 \sigma^{\lambda \rho} R_{\mu\nu\lambda\rho}$, where standard QED notations for the electromagnetic tensor and the spinor matrices are used.
We should emphasize that although CPT was developed for massless field theories, its generalization for massive theories like QED is simple. It boils down to an additional factor ${\mathrm e}^{- s m^2}$ under the proper time integral \cite{BarVilk-PRep85} that involves only form factors but not the curvatures. Explicit examples of such calculations can be seen in \cite{GShore-NPB02,GusZeln-PRD99}.

To look at the relation of the used techniques and their results 
to other approaches in quantum field theory,
we mention the so-called string inspired or world line methods
that were developed in recent years, e.g., see review \cite{Schubert-PR}.
Their core is the world line method \cite{Strass-NPB92}.
It is a path integral perturbation theory where the world line parameter is
effectively the Schwinger's proper time \cite{Schubert-PR}.
Indeed, as was shown in \cite{SchmSchub-PLB93}
the form factors of the one-loop effective action,
obtained by the world line method coincide with those
calculated in the covariant perturbation theory \cite{CPT2,BGVZ-NPB95}, in flat spacetime. Nevertheless, these approaches are different, and each has its own domain of applicability and the set of problem it is best suited for.

We expect that our results will be applied to gauge field theories.
The chiral perturbation theory was one of the applications of the Schwinger-DeWitt series \cite{Ball-PRep89}, so, more powerful techniques must find their use as well.
Still, the main motivation for the development of the covariant perturbation
theory was quantum gravity problems  \cite{Vilk-Strasb,Vilk-Gospel,Vilk-CQG92}.
Pursuing the systematic development of the effective action method for particle creation \cite{MirzVilk-AnnPh98}  G.A. Vilkovisky has solved the quantum gravitational collapse problem that appeared to have an intriguing and unexpected solution  \cite{Vilkov-PLB06}.

Further progress in the development of the covariant perturbation theory may lie in the construction of a partial summation of the curvature expansion and/or the derivation of the heat kernel  with separated points, $\hat{K}(s|x,y)$, but such developments are beyond the scope of this paper.

\section*{Acknowledgments}
\noindent
We acknowledge the support by NSERC of Canada at early stages of this work. 
The author is very grateful to A.O. Barvinsky and G.A. Vilkovisky for the long collaboration on the  the covariant perturbation theory.


\section*{Appendix A: The reduction equations for the form factors} \label{reduction}

This section reproduces the equations for the $\alpha$~-polynomial reduction of the form factors that were derived in \cite{CPT4,CPT-book}.

By eliminating all polynomials in $\alpha$~-parameters 
in the form factors of the heat kernel, the form factors 
can be explicitly expressed through the basic form factors,
\begin{eqnarray}
&&f(\xi)=\left\langle{\rm e}^{-\alpha_1\alpha_2\xi}\right\rangle_{2}
=\int^1_0\!d\alpha\,{\rm e}^{-\alpha(1-\alpha)\xi}, \label{fsmall} 
\\&&F(\xi_1,\xi_2,\xi_3)=\left\langle{\rm e}^{s\Omega}\right\rangle_{3}
= \nonumber\\&&\mbox{}
\int_{\alpha\geq 0}d^3\alpha\,
\delta(1-\alpha_1-\alpha_2-\alpha_3)
\exp(-\alpha_1\alpha_2\xi_3-\alpha_2\alpha_3\xi_1-
\alpha_1\alpha_3\xi_2).\label{Fbig} 
\end{eqnarray}
This kind of form factor representation is called, therefore, 
the explicit representation.

After the use of the delta-function in (\ref{fsmall}) and (\ref{Fbig}), 
there are two types of $\alpha$-monomials:
\begin{eqnarray}&&
\left\langle\alpha_1^n {\rm e}^{-\alpha_1\alpha_2\xi} \right\rangle_{2} =
\int^1_0 \! d\alpha\,\alpha^n \exp\Big[-\alpha(1-\alpha)\xi\Big],
\label{reduc-4}
\end{eqnarray}
\begin{eqnarray}&&
\left\langle\alpha_1^n \alpha_2^m {\rm e}^{s\Omega} \right\rangle_{3}=
\int^1_0 \! d\alpha_2
\int^{1-\alpha_2}_0 \! d\alpha_1 \,\alpha_1^n \alpha_2^m 
\nonumber\\&&\mbox{} \times
\exp\Big[-\alpha_2(1-\alpha_1-\alpha_2)\xi_1-
\alpha_1(1-\alpha_1-\alpha_2)\xi_2-\alpha_1\alpha_2\xi_3\Big].
\label{reduc-5}
\end{eqnarray}
For the case (\ref{reduc-4}) 
integration by parts produces
the following equation:
\begin{equation}
\int^1_0\!
d\alpha\,
\frac{d}{d\alpha}\alpha^n \exp\Big[-\alpha(1-\alpha)\xi\Big]=
\left\{
   \begin{array}{l}
          0,\hspace{7mm} n=0\\
          1,\hspace{7mm} n>0
       \end{array}
\right.  \label{reduc-6}
\end{equation}
which yields the recurrence relations,
\begin{eqnarray}&&
\left\langle\alpha_1 {\rm e}^{-\alpha_1\alpha_2\xi} \right\rangle_{2} =
\frac12 \left\langle {\rm e}^{-\alpha_1\alpha_2\xi} \right\rangle_{2},
\label{reduc-7}
\end{eqnarray}
\begin{eqnarray}&&
\left\langle\alpha_1^n {\rm e}^{-\alpha_1\alpha_2\xi} \right\rangle_{2} =
\frac12 \left\langle \alpha_1^{n-1}
{\rm e}^{-\alpha_1\alpha_2\xi}\right\rangle_{2}
\nonumber\\&&\ \ \ \ \ \ \ \ \mbox{}
-\frac12 (n-1)\left\langle\alpha_1^{n-2}
\left(\frac{{\rm e}^{-\alpha_1\alpha_2\xi}-1}{\xi}
\right)\right\rangle_{2},\hspace{7mm} n\geq 2.
\label{reduc-8}
\end{eqnarray}
They allow one to express all integrals (\ref{reduc-4}) through
the basic form factor (\ref{fsmall}).
Since the recurrence relations imply a division by $\xi$,
the appearing subtractions maintain the analyticity of 
the integral (\ref{reduc-4}) in $\xi$ at $\xi=0$.
For the form factor with subtractions in terms of (\ref{fsmall}) one has
\begin{eqnarray}&&
\left\langle \frac{{\rm e}^{-\alpha_1\alpha_2\xi}-1}{\xi}\right\rangle_{2}=
\frac{f(\xi)-1}{\xi}.  \label{reduc-11}
\end{eqnarray}

Elimination of the polynomials in $\alpha$ from  the
form factors   (\ref{reduc-5})  is based on integration by parts:
\begin{eqnarray}&&
\int^1_0 \! d\alpha_2 \int^{1-\alpha_2}_0 \! d\alpha_1
\frac{d}{d\alpha_1}\alpha_1^n \alpha_2^m \exp\left(\left.s\Omega
\right|_{\alpha_3=1-\alpha_1-\alpha_2}\right)
\nonumber\\&&\ \ \ \ \ \ \ \ \mbox{}=
\left\{
   \begin{array}{ll}
\left\langle\alpha_2^m\left({\rm e}^{-\alpha_1\alpha_2\xi_3}-
{\rm e}^{-\alpha_1\alpha_2\xi_1}\right)\right\rangle_{2},
&\hspace{7mm}
n=0
\\[2mm]
\left\langle\alpha_1^n\alpha_2^m {\rm e}^{-\alpha_1\alpha_2\xi_3}
\right\rangle_{2},&\hspace{7mm} n>0.
       \end{array}
\right.     \label{reduc-13}
\end{eqnarray}
A similar expression involving 
the differentiation with respect to $\alpha_2$ is
obtained from (\ref{reduc-13}) by transmutations of
the indices 1-2.
The first order form factors appearing on the right hand
sides of them are subject to the recurrence relations above.
By performing the differentiations on the left hand sides of
these two formulas, 
one obtains two linear algebraic equations for the quantities
$\left\langle\alpha_1^{n+1}\alpha_2^{m} {\rm e}^{s\Omega} 
\right\rangle_{3},
\hspace{7mm}
\left\langle\alpha_1^{n}\alpha_2^{m+1} {\rm e}^{s\Omega} 
\right\rangle_{3}
$containing the highest order monomials. The discriminant of this
linear system is
\begin{equation}
\Delta=\xi_1^2+\xi_2^2+\xi_3^2-2\xi_1\xi_2-2\xi_1\xi_3-2\xi_2\xi_3,
\label{reduc-15}
\end{equation}
and the recurrence relations obtained in this way are of the form
\begin{eqnarray}
\left\langle\alpha_1^{n+1}\alpha_2^{m} {\rm e}^{s\Omega} 
\right\rangle_{3}&=&
-\frac{\xi_1(\xi_3+\xi_2-\xi_1)}{\Delta}
\left\langle\alpha_1^{n}\alpha_2^{m} {\rm e}^{s\Omega} 
\right\rangle_{3}\nonumber\\&&\mbox{}
+2n\frac{\xi_1}{\Delta}
\left\langle\alpha_1^{n-1}\alpha_2^{m} {\rm e}^{s\Omega} 
\right\rangle_{3}\nonumber\\&&\mbox{}
+m\frac{(\xi_3-\xi_2-\xi_1)}{\Delta}
\left\langle\alpha_1^{n}\alpha_2^{m-1} {\rm e}^{s\Omega} 
\right\rangle_{3}\nonumber\\&&\mbox{}
-\frac{(\xi_3+\xi_1-\xi_2)}{\Delta}
\left\langle\alpha_1^{n}\alpha_2^{m} 
{\rm e}^{-\alpha_1\alpha_2\xi_3}\right\rangle_{2}
+\beta(n,m),    \label{reduc-16}
\end{eqnarray}
\begin{eqnarray}&&
\beta(n,m)=0,\hspace{7mm} n>0,\hspace{7mm} m>0   \label{reduc-17}
\end{eqnarray}
\begin{eqnarray}&&
\beta(n,0)=
\frac{(\xi_3-\xi_2-\xi_1)}{\Delta}
\left\langle\alpha_1^{n} {\rm e}^{-\alpha_1\alpha_2\xi_2}\right\rangle_{2},
\hspace{7mm} n>0   \label{reduc-18}
\end{eqnarray}
\begin{eqnarray}&&
\beta(0,m)=
2\frac{\xi_1}{\Delta}
\left\langle\alpha_1^{m} {\rm e}^{-\alpha_1\alpha_2\xi_1}\right\rangle_{2},
\hspace{7mm}
m>0    \label{reduc-19}
\end{eqnarray}
\begin{eqnarray}&&
\beta(0,0)=
2\frac{\xi_1}{\Delta}
\left\langle {\rm e}^{-\alpha_1\alpha_2\xi_1}\right\rangle_{2}
+\frac{(\xi_3-\xi_2-\xi_1)}{\Delta}
\left\langle {\rm e}^{-\alpha_1\alpha_2\xi_2}\right\rangle_{2}.
\label{reduc-20}
\end{eqnarray}
An expression for 
$ \left\langle\alpha_1^{n}\alpha_2^{m+1} 
{\rm e}^{s\Omega} \right\rangle_{3}$
can be obtained from 
(\ref{reduc-16})--(\ref{reduc-20}) by the 1-2 indexes transmutation.
Together with (\ref{reduc-7})--(\ref{reduc-8}) these relations make it possible
to express all integrals (\ref{reduc-5}) 
through the basic form factors $f(\xi)$
and $F(\xi_1,\xi_2,\xi_3)$. 
For the combination (\ref{alpha-12}) with a subtraction, one has
\begin{eqnarray}&&
\left\langle {\rm e}^{s\Omega}-1\right\rangle_{3}=
F(\xi_1,\xi_2,\xi_3) -\frac12,   \label{reduc-36}
\end{eqnarray}
in terms of (\ref{Fbig}). Analyticity in $\xi_1,\xi_2,\xi_3$
holds now only in the sum of the form factors on the
right hand side of (\ref{reduc-16}), 
and it is a nontrivial fact that, when these form factors
are expanded in power series in $\xi$, the denominator $\Delta$
gets always cancelled. The mechanism of maintaining
analyticity is based on the existence of linear differential
equations which the functions (\ref{fsmall}) and (\ref{Fbig}) satisfy.

The differential equations for the basic form factors can be derived
with the aid of the recurrence relations above. From (\ref{fsmall})
and (\ref{reduc-8}) one has the following equation for the function~\hbox{$f(\xi)$:}
\begin{eqnarray}
-\frac{d}{d\xi}f(\xi)&=&\frac14 f(\xi)+\frac12
\frac{f(\xi)-1}{\xi}.  \label{reduc-39}
\end{eqnarray}
Similarly, one obtains the equation for the form factor (\ref{Fbig}):
\begin{eqnarray}
-\frac{\partial}{\partial\xi_1}F(\xi_1,\xi_2,\xi_3)&=&
\frac1{{\Delta}^2}\Big[(\xi_1-\xi_2-\xi_3)\Delta\nonumber\\&&\mbox{}
+\xi_2\xi_3(2\xi_2\xi_3-\xi_2^2-\xi_3^2+\xi_1^2)\Big]F(\xi_1,\xi_2,\xi_3)
\nonumber\\&&\mbox{}
+\frac12\frac{8\xi_1\xi_2\xi_3+(\xi_2+\xi_3-\xi_1)\Delta}{\Delta^2}f(\xi_1)
\nonumber\\&&\mbox{}
+2\frac{\xi_2\xi_3(\xi_3-\xi_2-\xi_1)}{\Delta^2}f(\xi_2)\nonumber\\&&\mbox{}
+2\frac{\xi_2\xi_3(\xi_2-\xi_3-\xi_1)}{\Delta^2}f(\xi_3).  \label{reduc-41}
\end{eqnarray}
Finally, as a consequence of these equations, one can derive an
equation for the form factor (\ref{Fbig}) as a function of $s$:
\begin{eqnarray}
-s\frac{\partial}{\partial s}F(-s\Box_1,-s\Box_2,-s\Box_3) &=&
\left(s\frac{\Box_1\Box_2\Box_3}{D}+1\right)
F(-s\Box_1,-s\Box_2,-s\Box_3)\nonumber\\&&\mbox{}
+\frac{\Box_1(\Box_3+\Box_2-\Box_1)}{2D}f(-s\Box_1)\nonumber\\&&\mbox{}
+\frac{\Box_2(\Box_3+\Box_1-\Box_2)}{2D}f(-s\Box_2)\nonumber\\&&\mbox{}
+\frac{\Box_3(\Box_1+\Box_2-\Box_3)}{2D}f(-s\Box_3),   \label{reduc-42}
\end{eqnarray}
\begin{equation}
D={\Box_1}^2+{\Box_2}^2+{\Box_3}^2-2\Box_1\Box_2-2\Box_1\Box_3-2\Box_2\Box_3.
\label{reduc-43}
\end{equation}
This is an equation used for the study of a two-dimensional field model in section~\ref{Weyl}.


\section*{Appendix B: The explicit representation for the second order form factors} \label{finalK2}

The second order form factors of the heat kernel
are expressed through basic form factors (\ref{fsmall}), (\ref{Fbig})
and $\Delta$ (\ref{reduc-15}):

\begin{eqnarray}
&&G_1({\xi_1},{\xi_2},{\xi_3})= F({\xi_1},{\xi_2},{\xi_3}),   \label{G1}
\\[2mm]
&&G_2({\xi_1},{\xi_2},{\xi_3})= F({\xi_1},{\xi_2},{\xi_3})
\Big[ \frac{2 {\xi_1} {\xi_2}}{\Delta^2}
({\xi_3}+{\xi_2}-{\xi_1}) ({\xi_3}+{\xi_1}-{\xi_2})
\nonumber\\ &&\mbox{}
+ \frac{2}{\Delta}({\xi_3}-{\xi_2}-{\xi_1})\Big]
 - f({\xi_1})\frac{4 {\xi_1} {\xi_2}}{\Delta^2} ({\xi_3}+{\xi_1}-{\xi_2})
\nonumber\\ &&\mbox{}
 - f({\xi_2})\frac{4 {\xi_1} {\xi_2}}{\Delta^2} ({\xi_3}+{\xi_2}-{\xi_1})
\nonumber\\ &&\mbox{}
 -f({\xi_3})\frac{1}{\Delta^2}
(-6 {\xi_1} {\xi_2} {\xi_3}
-3 {\xi_1} {\xi_3}^2
\nonumber\\ &&\mbox{}
-3 {\xi_2} {\xi_3}^2
+3 {\xi_3} {\xi_1}^2
+3 {\xi_3} {\xi_2}^2
+{\xi_3}^3
+{\xi_1} {\xi_2}^2
+{\xi_2} {\xi_1}^2
-{\xi_2}^3
-{\xi_1}^3),                  \label{G2}
\\[2mm]
&&G_3({\xi_1},{\xi_2},{\xi_3})=
F({\xi_1},{\xi_2},{\xi_3})
\Big[ -\frac{1}{3 \Delta^2}
(-2 {\xi_1} {\xi_3} {\xi_2}^2
-2 {\xi_1} {\xi_2} {\xi_3}^2
\nonumber\\&&\mbox{}
+4 {\xi_1}^2 {\xi_2} {\xi_3}
+{\xi_1}^4
+{\xi_2}^4+{\xi_3}^4
-6 {\xi_1}^2 {\xi_3}^2
+2 {\xi_3} {\xi_1}^3
+2 {\xi_1} {\xi_3}^3
-6 {\xi_1}^2 {\xi_2}^2
\nonumber\\&&\mbox{}
+2 {\xi_2} {\xi_1}^3
+2 {\xi_1} {\xi_2}^3
+6 {\xi_2}^2 {\xi_3}^2
-4 {\xi_2}^3 {\xi_3}
-4 {\xi_2} {\xi_3}^3)
-\frac{4 {\xi_1}}{\Delta}\Big]
\nonumber\\&&\mbox{}
+f({\xi_1})\frac{4 {\xi_1}}{\Delta^2} 
(2 {\xi_2} {\xi_3}
-{\xi_3}^2
+{\xi_1} {\xi_3}
+{\xi_1} {\xi_2}
-{\xi_2}^2)
\nonumber\\&&\mbox{}
+f({\xi_2})\frac{1}{\Delta^2}({\xi_1}+{\xi_2}-{\xi_3}) 
({\xi_1}^2
+2 {\xi_2} {\xi_3}
-{\xi_2}^2
-{\xi_3}^2)
\nonumber\\&&\mbox{}
+f({\xi_3})\frac{1}{\Delta^2}
(2 {\xi_1} {\xi_2} {\xi_3}
-{\xi_1} {\xi_3}^2
+{\xi_3} {\xi_1}^2
+{\xi_2}^3
-{\xi_2} {\xi_1}^2
\nonumber\\&&\mbox{}
+{\xi_1}^3
-{\xi_1} {\xi_2}^2
+3 {\xi_2} {\xi_3}^2
-3 {\xi_3} {\xi_2}^2
-{\xi_3}^3),
\\[2mm]
&&G_4(\xi_1,{\xi_2},{\xi_3})=
F({\xi_1},{\xi_2},{\xi_3})\Big[
\frac{1}{36\Delta^4} 
({\xi_3}^8
-4 {\xi_1} {\xi_3}^7
\nonumber\\&&\mbox{}
-16 {\xi_1}^2 {\xi_3}^6
+68 {\xi_1}^3 {\xi_3}^5
-100 {\xi_1}^4 {\xi_3}^4
+68 {\xi_1}^5 {\xi_3}^3
-16 {\xi_1}^6 {\xi_3}^2
\nonumber\\&&\mbox{}
-4 {\xi_1}^7 {\xi_3}
+2 {\xi_1}^8
+16 {\xi_2} {\xi_1} {\xi_3}^6
-60 {\xi_2} {\xi_1}^2 {\xi_3}^5
+8 {\xi_2} {\xi_1}^3 {\xi_3}^4
\nonumber\\&&\mbox{}
+68 {\xi_2} {\xi_1}^4 {\xi_3}^3
-48 {\xi_2} {\xi_3}^2 {\xi_1}^5
-4 {\xi_2} {\xi_3} {\xi_1}^6
+8 {\xi_2} {\xi_1}^7
+96 {\xi_1}^2 {\xi_2}^2 {\xi_3}^4
\nonumber\\&&\mbox{}
-136 {\xi_1}^3 {\xi_2}^2 {\xi_3}^3
+36 {\xi_1}^5 {\xi_2}^2 {\xi_3}
-16 {\xi_1}^6 {\xi_2}^2
+64 {\xi_1}^3 {\xi_2}^3 {\xi_3}^2
-28 {\xi_3} {\xi_1}^4 {\xi_2}^3
\nonumber\\&&\mbox{}
-40 {\xi_1}^5 {\xi_2}^3
+46 {\xi_1}^4 {\xi_2}^4)
\nonumber\\&&\mbox{}
 +\frac{1}{3\Delta^3}
 (3 {\xi_3}^5-16 {\xi_1} {\xi_3}^4+4 {\xi_1}^2 {\xi_3}^3
\nonumber\\&&\mbox{}
+24 {\xi_3}^2 {\xi_1}^3
-26 {\xi_3} {\xi_1}^4
+8 {\xi_1}^5+
28 {\xi_1} {\xi_3}^3 {\xi_2}
-52 {\xi_3}^2 {\xi_1}^2 {\xi_2}
\nonumber\\&&\mbox{}
+12 {\xi_1}^4 {\xi_2}
+26 {\xi_1}^2 {\xi_3} {\xi_2}^2
-20 {\xi_1}^3 {\xi_2}^2)\Big]
\nonumber\\&&\mbox{}
+\Big(F({\xi_1},{\xi_2},{\xi_3})-\frac12\Big)
\frac{2}{\Delta^2} 
({\xi_3}^2-4 {\xi_1} {\xi_3}
+2 {\xi_1}^2+4 {\xi_1} {\xi_2})
\nonumber\\&&\mbox{}
+f({\xi_1}) \frac{1}{6\Delta^4} 
(-{\xi_1}^2 {\xi_3}^5
+{\xi_3}^2 {\xi_1}^5
+{\xi_1}^6 {\xi_2}
\nonumber\\&&\mbox{}
-{\xi_1} {\xi_2}^6
-{\xi_2}^7+{\xi_1}^7
+10 {\xi_3} {\xi_1}^5 {\xi_2}
+5 {\xi_2}^4 {\xi_3}^3
-{\xi_3}^7
-9 {\xi_2}^5 {\xi_3}^2
\nonumber\\&&\mbox{}
+5 {\xi_3} {\xi_2}^6
+5 {\xi_2} {\xi_3}^6
+23 {\xi_1}^2 {\xi_3}^4 {\xi_2}
+57 {\xi_1} {\xi_3}^4 {\xi_2}^2
+12 {\xi_1}^3 {\xi_3}^3 {\xi_2}
\nonumber\\&&\mbox{}
-29 {\xi_3}^2 {\xi_1}^4 {\xi_2}
-20 {\xi_1}^3 {\xi_3} {\xi_2}^3
-6 {\xi_1} {\xi_3} {\xi_2}^5
+37 {\xi_1} {\xi_3}^2 {\xi_2}^4
\nonumber\\&&\mbox{}
+11 {\xi_1}^2 {\xi_3} {\xi_2}^4
+3 {\xi_1} {\xi_3}^6
-42 {\xi_1}^2 {\xi_3}^3 {\xi_2}^2
-68 {\xi_1} {\xi_3}^3 {\xi_2}^3
-22 {\xi_1} {\xi_3}^5 {\xi_2}
\nonumber\\&&\mbox{}
+14 {\xi_1}^2 {\xi_3}^2 {\xi_2}^3
-14 {\xi_3}^2 {\xi_1}^3 {\xi_2}^2
+3 {\xi_3} {\xi_1}^4 {\xi_2}^2
-5 {\xi_1}^3 {\xi_3}^4
+5 {\xi_1}^4 {\xi_3}^3
\nonumber\\&&\mbox{}
-3 {\xi_3} {\xi_1}^6
+5 {\xi_1}^5 {\xi_2}^2
-9 {\xi_3}^5 {\xi_2}^2
-27 {\xi_1}^4 {\xi_2}^3
+27 {\xi_1}^3 {\xi_2}^4
\nonumber\\&&\mbox{}
-5 {\xi_1}^2 {\xi_2}^5
+5 {\xi_2}^3 {\xi_3}^4)
\nonumber\\&&\mbox{}
-f({\xi_3})\frac{1}{24\Delta^4}
 (-{\xi_3}^7
-2 {\xi_1} {\xi_3}^6
+38 {\xi_1}^2 {\xi_3}^5
-90 {\xi_1}^3 {\xi_3}^4
\nonumber\\&&\mbox{}
+90 {\xi_1}^4 {\xi_3}^3
-38 {\xi_3}^2 {\xi_1}^5
+2 {\xi_3} {\xi_1}^6
+2 {\xi_1}^7
-10 {\xi_1} {\xi_3}^5 {\xi_2}
+50 {\xi_1}^2 {\xi_3}^4 {\xi_2}
\nonumber\\&&\mbox{}
-24 {\xi_1}^3 {\xi_3}^3 {\xi_2}
+2 {\xi_3}^2 {\xi_1}^4 {\xi_2}
-20 {\xi_3} {\xi_1}^5 {\xi_2}
+14 {\xi_1}^6 {\xi_2}
-66 {\xi_1}^2 {\xi_3}^3 {\xi_2}^2
\nonumber\\&&\mbox{}
+36 {\xi_3}^2 {\xi_1}^3 {\xi_2}^2
+62 {\xi_3} {\xi_1}^4 {\xi_2}^2
-54 {\xi_1}^5 {\xi_2}^2
-44 {\xi_1}^3 {\xi_3} {\xi_2}^3
+38 {\xi_1}^4 {\xi_2}^3)
\nonumber\\&&\mbox{}
+\left(\frac{f({\xi_1})-1}{{\xi_1}}\right)
\frac{ 2 {\xi_1}}{\Delta^3} 
(4 {\xi_1}^3 {\xi_2}
+3 {\xi_1}^4
+2 {\xi_1}^2 {\xi_2}^2
-8 {\xi_1} {\xi_2}^3
\nonumber\\&&\mbox{}
-{\xi_2}^4
-8 {\xi_3} {\xi_1}^3
+4 {\xi_3} {\xi_2}^3
+16 {\xi_1} {\xi_3} {\xi_2}^2
-6 {\xi_2}^2 {\xi_3}^2
+6 {\xi_1}^2 {\xi_3}^2
\nonumber\\&&\mbox{}
-{\xi_3}^4
-8 {\xi_1} {\xi_2} {\xi_3}^2
+4 {\xi_2} {\xi_3}^3)
\nonumber\\&&\mbox{}
+\left(\frac{f({\xi_3})-1}{{\xi_3}}\right)
\frac{ 1}{4\Delta^3} 
(7 {\xi_3}^5
-22 {\xi_1} {\xi_3}^4
-20 {\xi_1}^2 {\xi_3}^3
\nonumber\\&&\mbox{}
+52 {\xi_3}^2 {\xi_1}^3
-26 {\xi_3} {\xi_1}^4
+2 {\xi_1}^5
+36 {\xi_1} {\xi_3}^3 {\xi_2}
-52 {\xi_3}^2 {\xi_1}^2 {\xi_2}
\nonumber\\&&\mbox{}
+8 {\xi_3} {\xi_1}^3 {\xi_2}
-6 {\xi_1}^4 {\xi_2}
+18 {\xi_1}^2 {\xi_3} {\xi_2}^2
+4 {\xi_1}^3 {\xi_2}^2),
\\[2mm]
&&G_5({\xi_1},{\xi_2},{\xi_3})
=\left(F({\xi_1},{\xi_2},{\xi_3})-\frac12\right)
{2\over {{\xi_1} {\xi_2}}}
-f({\xi_3}){{(-2 {\xi_2} + {\xi_3})}\over {8 {\xi_1} {\xi_2}}}
\nonumber\\&&\mbox{}
-\left(\frac{f({\xi_3})-1}{{\xi_3}}\right)
{{(-2 {\xi_2} + 5 {\xi_3})}\over {4 {\xi_1} {\xi_2}}},
\\[2mm]
&&G_6({\xi_1},{\xi_2},{\xi_3})
=F({\xi_1},{\xi_2},{\xi_3})\Big[
-\frac{2}{{\Delta}^{2}}( -{\xi_1} + {\xi_2} - {\xi_3} )
\times\nonumber\\&&\mbox{}\times
  ( -{\xi_1} - {\xi_2} + {\xi_3} )
  ( -{\xi_1} + {\xi_2} + {\xi_3} )
+{8\over {\Delta}}\Big]
\nonumber\\&&\mbox{}
+f({\xi_1})\frac{4}{{\Delta}^{2}}
( -{\xi_1}
+{\xi_2} 
- {\xi_3} )  
( -{\xi_1} 
- {\xi_2} 
+ {\xi_3} )
\nonumber\\&&\mbox{}
-f({\xi_2})\frac{4}{{\Delta}^{2}}
( -{\xi_1} - {\xi_2} + {\xi_3} )
 ( -{\xi_1} + {\xi_2} + {\xi_3} )
\nonumber\\&&\mbox{}
-f({\xi_3})\frac{4}{{\Delta}^{2}}
( {{{\xi_1}}^2}- 2 {\xi_1} {\xi_2} + {{{\xi_2}}^2} - {{{\xi_3}}^2} ),
\\[2mm]
&&G_7({\xi_1},{\xi_2},{\xi_3})=
F({\xi_1},{\xi_2},{\xi_3})
 \frac{2}{\Delta}
(-{\xi_3}-{\xi_1}+{\xi_2})
-f({\xi_1})\frac{2}{{\xi_2}\Delta}
(-{\xi_3}+{\xi_2}+{\xi_1})
\nonumber\\&&\mbox{}
+f({\xi_2})\frac{4}{\Delta}
-f({\xi_3}) \frac{2}{{\xi_2} \Delta}
({\xi_3}+{\xi_2}-{\xi_1})
\nonumber\\&&\mbox{}
+\frac{1}{({\xi_1}-{\xi_3})}
\Big(f({\xi_1})-f({\xi_3})\Big)
\frac{2}{{\xi_2}},
\\[2mm]
&&G_8({\xi_1},{\xi_2},{\xi_3})=
F({\xi_1},{\xi_2},{\xi_3})
\Big[
-\frac{ 4 {\xi_2}}{ \Delta^2}
(-{\xi_3}-{\xi_1}+{\xi_2})^2   
 -\frac{8}{\Delta}
\Big]
\nonumber\\&&\mbox{}
+f({\xi_1})\frac{2}{{\xi_2}\Delta^2}
(-{\xi_3}+{\xi_2}+{\xi_1}) 
(-4 {\xi_2} {\xi_3}-4 {\xi_1} {\xi_2}
+3 {\xi_2}^2
-2 {\xi_1} {\xi_3}
+{\xi_1}^2
+{\xi_3}^2)
\nonumber\\&&\mbox{}
+f({\xi_2})\frac{ 8 {\xi_2}}{\Delta^2} 
({\xi_3}+{\xi_1}-{\xi_2})
\nonumber\\&&\mbox{}
+f({\xi_3})
\frac{2}{{\xi_2}\Delta^2}
({\xi_3}+{\xi_2}-{\xi_1})
(-4 {\xi_2} {\xi_3}
-4 {\xi_1} {\xi_2}
+3 {\xi_2}^2
-2 {\xi_1} {\xi_3}
+{\xi_1}^2+{\xi_3}^2)
\nonumber\\&&\mbox{}
-\frac{1}{({\xi_1}-{\xi_3})}
\Big(f({\xi_1})-f({\xi_3})\Big)
\frac{2}{{\xi_2}},
\\[2mm]
&&G_9({\xi_1},{\xi_2},{\xi_3})
=F({\xi_1},{\xi_2},{\xi_3})\frac{8}{{\Delta}^{2}}
( -2 {{{\xi_1}}^2} 
+ 2 {\xi_1} {\xi_2} 
+ {{{\xi_3}}^2} )
\nonumber\\&&\mbox{}
-\Big(F({\xi_1},{\xi_2},{\xi_3})-\frac12\Big){8\over {\Delta {\xi_1} {\xi_2}}}
( 2 {\xi_1} - {\xi_3} )
\nonumber\\&&\mbox{}
-f({\xi_3})
{1\over {2 {\xi_1} {\xi_2}}}
+\frac{f({{\xi_1}})-1}{{\xi_1}}
\frac{32 {\xi_1}}{{\Delta}^{2}}
 ( -{\xi_1} + {\xi_2} - {\xi_3} )
\nonumber\\&&\mbox{}
-\frac{f({\xi_3})-1}{{\xi_3}}
{1\over {{{\Delta}^2} {\xi_1} {\xi_2}}}
( 2 {{{\xi_1}}^4} 
- 8 {{{\xi_1}}^3} {\xi_2} 
+ 6 {{{\xi_1}}^2} {{{\xi_2}}^2} -
    16 {{{\xi_1}}^3} {\xi_3} 
\nonumber\\&&\mbox{}
+ 16 {{{\xi_1}}^2} {\xi_2} {\xi_3} 
+ 36 {{{\xi_1}}^2} {{{\xi_3}}^2} 
- 20 {\xi_1} {\xi_2} {{{\xi_3}}^2} 
- 32 {\xi_1} {{{\xi_3}}^3} 
+ 5 {{{\xi_3}}^4} ),
\\[2mm]
&&G_{10}({\xi_1},{\xi_2},{\xi_3})=
F({\xi_1},{\xi_2},{\xi_3})
\Big[
\frac{ 2}{3 {\xi_2} \Delta^4}
({\xi_3}+{\xi_1}-{\xi_2})^2 
(-2 {\xi_2} {\xi_3} {\xi_1}^2
\nonumber\\&&\mbox{}
-2 {\xi_2} {\xi_1} {\xi_3}^2
+4 {\xi_1} {\xi_2}^2 {\xi_3}
+{\xi_2}^4
+{\xi_1}^4
+{\xi_3}^4
-6 {\xi_2}^2 {\xi_3}^2
+2 {\xi_3} {\xi_2}^3
+2 {\xi_2} {\xi_3}^3
\nonumber\\&&\mbox{}
-6 {\xi_1}^2 {\xi_2}^2
+2 {\xi_2} {\xi_1}^3
+2 {\xi_1} {\xi_2}^3
+6 {\xi_1}^2 {\xi_3}^2
-4 {\xi_1}^3 {\xi_3}
-4 {\xi_1} {\xi_3}^3)
\nonumber\\&&\mbox{}
+\frac{4}{3 \Delta^3}
(-14 {\xi_2} {\xi_3} {\xi_1}^2
+14 {\xi_2} {\xi_3}^3
+40 {\xi_1} {\xi_2}^2 {\xi_3}
-14 {\xi_2} {\xi_1} {\xi_3}^2
\nonumber\\&&\mbox{}
-22 {\xi_3} {\xi_2}^3
-12 {\xi_2}^2 {\xi_3}^2
-4 {\xi_1}^3 {\xi_3}
-4 {\xi_1} {\xi_3}^3
+6 {\xi_1}^2 {\xi_3}^2
+{\xi_3}^4
\nonumber\\&&\mbox{}
+19 {\xi_2}^4
-12 {\xi_1}^2 {\xi_2}^2
+14 {\xi_2} {\xi_1}^3
-22 {\xi_1} {\xi_2}^3
+{\xi_1}^4)
\Big]
\nonumber\\&&\mbox{}
+\left(F({\xi_1},{\xi_2},{\xi_3})-\frac12\right)
\frac{48 {\xi_2}}{\Delta^2}
\nonumber\\&&\mbox{}
+f({\xi_1})
\frac{1}{6{\xi_2} \Delta^4} 
({\xi_2}+{\xi_1}-{\xi_3}) 
(-16 {\xi_2} {\xi_1}^2 {\xi_3}^3
-20 {\xi_3} {\xi_2}^2 {\xi_1}^3
+24 {\xi_2} {\xi_3} {\xi_1}^4
\nonumber\\&&\mbox{}
+8 {\xi_3} {\xi_2}^5
-64 {\xi_2}^3 {\xi_3}^3
+35 {\xi_3}^2 {\xi_2}^4
+37 {\xi_2}^2 {\xi_3}^4
+8 {\xi_1} {\xi_2}^5
-64 {\xi_1}^3 {\xi_2}^3
\nonumber\\&&\mbox{}
+37 {\xi_2}^2 {\xi_1}^4
+35 {\xi_1}^2 {\xi_2}^4
-8 {\xi_2} {\xi_3}^5
-8 {\xi_2} {\xi_1}^5
-9 {\xi_2}^6
+{\xi_1}^6
+{\xi_3}^6
\nonumber\\&&\mbox{}
-20 {\xi_2}^2 {\xi_1} {\xi_3}^3
-34 {\xi_2}^2 {\xi_3}^2 {\xi_1}^2
-6 {\xi_3} {\xi_1} {\xi_2}^4
-16 {\xi_2} {\xi_1}^3 {\xi_3}^2
+24 {\xi_2} {\xi_1} {\xi_3}^4
\nonumber\\&&\mbox{}
-6 {\xi_1}^5 {\xi_3}
-6 {\xi_1} {\xi_3}^5
-20 {\xi_1}^3 {\xi_3}^3
+15 {\xi_1}^4 {\xi_3}^2
+15 {\xi_1}^2 {\xi_3}^4)
\nonumber\\&&\mbox{}
-f({\xi_2})
\frac{4 {\xi_2}}{3 \Delta^4}
({\xi_3}+{\xi_1}-{\xi_2}) 
(-2 {\xi_2} {\xi_3} {\xi_1}^2
-2 {\xi_2} {\xi_1} {\xi_3}^2
+4 {\xi_1} {\xi_2}^2 {\xi_3}
\nonumber\\&&\mbox{}
+{\xi_2}^4
+{\xi_1}^4
+{\xi_3}^4
-6 {\xi_2}^2 {\xi_3}^2
+2 {\xi_3} {\xi_2}^3
+2 {\xi_2} {\xi_3}^3
-6 {\xi_1}^2 {\xi_2}^2
\nonumber\\&&\mbox{}
+2 {\xi_2} {\xi_1}^3
+2 {\xi_1} {\xi_2}^3
+6 {\xi_1}^2 {\xi_3}^2
-4 {\xi_1}^3 {\xi_3}
-4 {\xi_1} {\xi_3}^3)
\nonumber\\&&\mbox{}
-f({\xi_3})
\frac{1}{6 {\xi_2} \Delta^4}
(-41 {\xi_3} {\xi_1}^2 {\xi_2}^4
+34 {\xi_3}^2 {\xi_1}^2 {\xi_2}^3
+30 {\xi_2}^2 {\xi_3}^3 {\xi_1}^2
+41 {\xi_3}^2 {\xi_1} {\xi_2}^4
\nonumber\\&&\mbox{}
-44 {\xi_1} {\xi_2}^3 {\xi_3}^3
+20 {\xi_2} {\xi_1}^3 {\xi_3}^3
-26 {\xi_2} {\xi_1} {\xi_3}^5
+9 {\xi_2}^7+{\xi_1}^7
+84 {\xi_3} {\xi_1}^3 {\xi_2}^3
\nonumber\\&&\mbox{}
-81 {\xi_3} {\xi_2}^2 {\xi_1}^4
+38 {\xi_3} {\xi_2} {\xi_1}^5
+7 {\xi_2} {\xi_3}^6
+33 {\xi_2}^2 {\xi_1} {\xi_3}^4
+25 {\xi_2} {\xi_1}^2 {\xi_3}^4
\nonumber\\&&\mbox{}
+2 {\xi_3}^2 {\xi_2}^2 {\xi_1}^3
-55 {\xi_2} {\xi_3}^2 {\xi_1}^4
-43 {\xi_3}^2 {\xi_2}^5
+27 {\xi_2}^3 {\xi_3}^4
+29 {\xi_3}^3 {\xi_2}^4
-29 {\xi_2}^2 {\xi_3}^5
\nonumber\\&&\mbox{}
-{\xi_3}^7
+6 {\xi_3} {\xi_1} {\xi_2}^5
+{\xi_3} {\xi_2}^6
-7 {\xi_3} {\xi_1}^6
+21 {\xi_1}^5 {\xi_3}^2
+7 {\xi_1} {\xi_3}^6
\nonumber\\&&\mbox{}
+35 {\xi_1}^3 {\xi_3}^4
-35 {\xi_1}^4 {\xi_3}^3
-21 {\xi_1}^2 {\xi_3}^5
-17 {\xi_1} {\xi_2}^6
+99 {\xi_1}^3 {\xi_2}^4
-101 {\xi_2}^3 {\xi_1}^4
\nonumber\\&&\mbox{}
-27 {\xi_1}^2 {\xi_2}^5
+45 {\xi_2}^2 {\xi_1}^5
-9 {\xi_2} {\xi_1}^6)
\nonumber\\&&\mbox{}
+\left(\frac{f({\xi_1})-1}{{\xi_1}}\right)
\frac{1}{{\xi_2} \Delta^3}
({\xi_2}+{\xi_1}-{\xi_3}) 
(-2 {\xi_2} {\xi_3} {\xi_1}^2
+10 {\xi_2} {\xi_1} {\xi_3}^2
\nonumber\\&&\mbox{}
+40 {\xi_1} {\xi_2}^2 {\xi_3}
+3 {\xi_2}^4
+{\xi_1}^4
+{\xi_3}^4
+12 {\xi_2}^2 {\xi_3}^2
-10 {\xi_3} {\xi_2}^3
-6 {\xi_2} {\xi_3}^3
\nonumber\\&&\mbox{}
+44 {\xi_1}^2 {\xi_2}^2
-2 {\xi_2} {\xi_1}^3
-46 {\xi_1} {\xi_2}^3
+6 {\xi_1}^2 {\xi_3}^2
-4 {\xi_1}^3 {\xi_3}
-4 {\xi_1} {\xi_3}^3)
\nonumber\\&&\mbox{}
-\left(\frac{f({\xi_2})-1}{{\xi_2}}\right)
\frac{16 {\xi_2}^2}{ \Delta^3}
(-4 {\xi_1} {\xi_3}
+2 {\xi_3}^2
+{\xi_2} {\xi_3}
+{\xi_1} {\xi_2}
+2 {\xi_1}^2
-3 {\xi_2}^2)
\nonumber\\&&\mbox{}
-\left(\frac{f({\xi_3})-1}{{\xi_3}}\right)
\frac{1}{{\xi_2}\Delta^3}
(4 {\xi_2} {\xi_1} {\xi_3}^3
+18 {\xi_3} {\xi_1}^2 {\xi_2}^2
+20 {\xi_3} {\xi_2} {\xi_1}^3
-3 {\xi_2}^5
\nonumber\\&&\mbox{}
-22 {\xi_1}^2 {\xi_2}^3
+18 {\xi_1}^3 {\xi_2}^2
+13 {\xi_1} {\xi_2}^4
-7 {\xi_2} {\xi_1}^4
+{\xi_1}^5
-42 {\xi_2}^2 {\xi_3}^3
\nonumber\\&&\mbox{}
-76 {\xi_3} {\xi_1} {\xi_2}^3
+6 {\xi_1} {\xi_2}^2 {\xi_3}^2
+43 {\xi_3} {\xi_2}^4
+2 {\xi_3}^2 {\xi_2}^3
-18 {\xi_2} {\xi_3}^2 {\xi_1}^2
+{\xi_2} {\xi_3}^4
\nonumber\\&&\mbox{}
-5 {\xi_3} {\xi_1}^4
-{\xi_3}^5
-10 {\xi_1}^2 {\xi_3}^3
+10 {\xi_1}^3 {\xi_3}^2
+5 {\xi_1} {\xi_3}^4)
\nonumber\\&&\mbox{}
-\frac{1}{({\xi_1}-{\xi_3})}
\Big(f({\xi_1})-f({\xi_3})\Big)
 \frac{1}{6{\xi_2}}
\nonumber\\&&\mbox{}
-\frac{1}{({\xi_1}-{\xi_3})}
\left(\frac{f({\xi_1})-1}{{\xi_1}}-\frac{f({\xi_3})-1}{{\xi_3}}\right)
\frac{ 1}{{\xi_2}},
\\[2mm]
&&G_{11}({\xi_1},{\xi_2},{\xi_3})
=F({\xi_1},{\xi_2},{\xi_3})
\Big[\frac{4 {\xi_1} {\xi_2}}{{\Delta}^{4}}
 ( 2 {{{\xi_1}}^4} 
- 8 {{{\xi_1}}^3} {\xi_2} 
+ 6 {{{\xi_1}}^2} {{{\xi_2}}^2} 
-    4 {{{\xi_1}}^2} {{{\xi_3}}^2} 
\nonumber\\&&\mbox{}
+ 4 {\xi_1} {\xi_2} {{{\xi_3}}^2} 
+ {{{\xi_3}}^4} )
-\frac{16}{{\Delta}^{3}}
( -3 {{{\xi_1}}^3} 
+ 3 {{{\xi_1}}^2} {\xi_2} 
+ 4 {{{\xi_1}}^2} {\xi_3}
\nonumber\\&&\mbox{}
    - 4 {\xi_1} {\xi_2} {\xi_3} 
+  {\xi_1} {{{\xi_3}}^2} 
- {{{\xi_3}}^3} )\Big]
\nonumber\\&&\mbox{}
+\Big(F({\xi_1},{\xi_2},{\xi_3})-\frac12\Big)
{8\over {{{\Delta}^2} {\xi_1} {\xi_2}}}
( 2 {{{\xi_1}}^2} 
+ 4 {\xi_1} {\xi_2}
- 4 {\xi_1} {\xi_3} 
+ {{{\xi_3}}^2} )
\nonumber\\&&\mbox{}
-f({\xi_1})\frac{16 {\xi_1} {\xi_2}}{{\Delta}^{4}}
 ( -{\xi_1} + {\xi_2} - {\xi_3} )^2
  ( -{\xi_1} + {\xi_2} + {\xi_3} )
\nonumber\\&&\mbox{}
-f({\xi_3}){1\over {2 {{\Delta}^4} {\xi_1} {\xi_2}}}
( -2 {{{\xi_1}}^7} 
+ 18 {{{\xi_1}}^6} {\xi_2} 
- 42 {{{\xi_1}}^5} {{{\xi_2}}^2} 
+    26 {{{\xi_1}}^4} {{{\xi_2}}^3} 
\nonumber\\&&\mbox{}
+ 14 {{{\xi_1}}^6} {\xi_3} 
- 76 {{{\xi_1}}^5} {\xi_2} {\xi_3} 
+    178 {{{\xi_1}}^4} {{{\xi_2}}^2} {\xi_3} 
- 116 {{{\xi_1}}^3} {{{\xi_2}}^3} {\xi_3} 
\nonumber\\&&\mbox{}
-    42 {{{\xi_1}}^5} {{{\xi_3}}^2} 
+ 110 {{{\xi_1}}^4} {\xi_2} {{{\xi_3}}^2} 
-    68 {{{\xi_1}}^3} {{{\xi_2}}^2} {{{\xi_3}}^2} 
+ 70 {{{\xi_1}}^4} {{{\xi_3}}^3}
 \nonumber\\&&\mbox{}
-    40 {{{\xi_1}}^3} {\xi_2} {{{\xi_3}}^3} 
- 30 {{{\xi_1}}^2} {{{\xi_2}}^2} {{{\xi_3}}^3} 
-    70 {{{\xi_1}}^3} {{{\xi_3}}^4} 
- 50 {{{\xi_1}}^2} {\xi_2} {{{\xi_3}}^4}
\nonumber\\&&\mbox{} 
+    42 {{{\xi_1}}^2} {{{\xi_3}}^5} 
+ 26 {\xi_1} {\xi_2} {{{\xi_3}}^5} 
- 14 {\xi_1} {{{\xi_3}}^6} 
+    {{{\xi_3}}^7} )
\nonumber\\&&\mbox{}
+\left(\frac{f({{\xi_1}})-1}{{\xi_1}}\right)
\frac{32 {\xi_1}}{{\Delta}^{3}}
 ( 3 {{{\xi_1}}^2} - {\xi_1} {\xi_2}
- 2 {{{\xi_2}}^2} - {\xi_1} {\xi_3} 
+ 4 {\xi_2} {\xi_3} - 2 {{{\xi_3}}^2} )
\nonumber\\&&\mbox{}
-\left(\frac{f({{\xi_3}})-1}{{\xi_3}}\right)
{1\over {{{\Delta}^3} {\xi_1} {\xi_2}}}
( -2 {{{\xi_1}}^5} 
+ 6 {{{\xi_1}}^4} {\xi_2} 
- 4 {{{\xi_1}}^3} {{{\xi_2}}^2} 
+    18 {{{\xi_1}}^4} {\xi_3} 
\nonumber\\&&\mbox{}
+ 24 {{{\xi_1}}^3} {\xi_2} {\xi_3} 
- 42 {{{\xi_1}}^2} {{{\xi_2}}^2} {\xi_3} 
-    52 {{{\xi_1}}^3} {{{\xi_3}}^2} 
+ 52 {{{\xi_1}}^2} {\xi_2} {{{\xi_3}}^2} 
\nonumber\\&&\mbox{}
+    68 {{{\xi_1}}^2} {{{\xi_3}}^3} 
- 20 {\xi_1} {\xi_2} {{{\xi_3}}^3} 
- 42 {\xi_1} {{{\xi_3}}^4} 
+     5{{{\xi_3}}^5} ),       \label{G11}
\end{eqnarray}


\end{document}